\documentclass{aa}  

\usepackage{graphicx}
\usepackage{txfonts,textcomp}
\usepackage{hyperref}
\newcommand{\chandra}{{\it Chandra}}
\newcommand{\swift}{{\it Swift}}
\newcommand{\xmm}{{\it XMM-Newton}}

\begin{document} 

\title{High-resolution X-ray spectra of RS\,Ophiuchi (2006 and 2021): Revealing the cause of SSS variability}
%\title{The disparity between the super-soft source phases of the 2006 and 2021 eruptions of RS Ophiuchi was due to absorption}
%\title{Solution to the disparity of SSS brightness of the 2006 and 2021 eruptions of RS Ophiuchi}
%\title{New insights from spectral variability studies of the SSS phases of the 2006 and 2021 eruptions of RS\,Ophiuchi}

   \author{J.-U. Ness\inst{\ref{esa}}\and
  A.P. Beardmore\inst{\ref{leicester}}\and
  M.F. Bode\inst{\ref{liverpool},\ref{botswana}}\and
  M.J. Darnley\inst{\ref{liverpool}}\and
  A. Dobrotka\inst{\ref{slovak}}\and
  J.J. Drake\inst{\ref{sao}}\and
  J. Magdolen\inst{\ref{slovak}}\and
  U. Munari\inst{\ref{asiago}}\and
  J.P. Osborne\inst{\ref{leicester}}\and
  M. Orio\inst{\ref{wisconsin},\ref{padova}}\and
  K.L. Page\inst{\ref{leicester}}\and
  S. Starrfield\inst{\ref{asu}}
    }

\institute{European Space Agency (ESA), European Space Astronomy Centre (ESAC), Camino Bajo del Castillo s/n, 28692 Villanueva de la Ca\~nada, Madrid, Spain; corresponding author: \email{jan.uwe.ness@esa.int}\label{esa}
\and
School of Physics \& Astronomy, University of Leicester, Leicester, LE1 7RH, UK\label{leicester}
        \and
Astrophysics Research Institute, Liverpool John Moores University
, IC2 Liverpool Science Park, Liverpool L3 5RF, UK\label{liverpool}
        \and
        Vice Chancellor's Office, Botswana International University of Science and Technology, Private Bag 16, Palapye, Botswana\label{botswana}
        \and
Advanced Technologies Research Institute, Faculty of Materials Science and Technology in Trnava, Slovak University of Technology in Bratislava, Bottova 25, 917 24 Trnava, Slovakia\label{slovak}
        \and
        Harvard-Smithsonian Center for Astrophysics, 60 Garden Street, Cambridge, MA 02138, USA\label{sao}
        \and
            %Osservatorio astronomico di Padova, Via Osservatorio 8, Asiago, Italy\label{asiago}
            INAF Astronomical Observatory of Padova, 36012 Asiago (VI), Italy\label{asiago}
\and
Dept. of Astronomy, University of Wisconsin, 475 N, Charter Str., Madison, WI 53706, USA\label{wisconsin}
     \and
INAF-Osservatorio di Padova, Vicolo Osservatorio 5, 35122 Padova, Italy\label{padova}
      \and
    School of Earth and Space Exploration, Arizona State University, Tempe, AZ 85287-1404, USA\label{asu}
}
   \authorrunning{Ness et al.}
   \titlerunning{2021 outburst of RS\,Oph in X-rays}
   \date{Received \today; accepted }

% \abstract{}{}{}{}{} 
% 5 {} token are mandatory
 
  \abstract
% Context
{
The $\sim 10-20$ year recurrent symbiotic nova RS\,Oph exploded on 2021 August 9,
the seventh
 confirmed recorded outburst since 1898. During the previous outburst
in 2006, the current fleet of X-ray space observatories was already in operation,
and thanks to the longevity of \swift, \xmm, and \chandra, a direct
comparison between these two outbursts is possible.
The \swift\ monitoring campaign revealed similar behaviour
during the early shock phase but very different behaviour during the super-soft source (SSS) phase. Two \xmm\thanks{\xmm\ is an ESA science mission with instruments and contributions directly funded by ESA Member States and NASA.}
observations were made during the 2021 SSS phase on days 37.1 and
55.6 after the 2021 optical peak. We focus in this work on the bright
SSS observation on day 55.6 and compare to SSS \chandra\ and \xmm\
grating observations made
on days 39.7, 54, and 66.9 after the 2006 optical peak.
}
% aims heading (mandatory)
{
By exploring the reasons for the differences between the 2006 and 2021
outbursts, we aim to obtain a better general understanding of the emission
and absorption mechanisms. While the emission mechanisms hold the key to
the physics of novae and nuclear burning, absorption processes may dominate
what we observe, and we aim to explore the cause of the gross initial variability
in the observed SSS emission.
}
  % methods heading (mandatory)
{
We present a novel approach to down-scaling the observed (brighter) 2006
SSS spectra to match the 2021 day 55.6 spectrum by parameter optimisation
of: (1) a constant factor (representing fainter source emission, smaller radius,
eclipses, etc.),
(2) a multi-ionisation photoelectric absorption model (representing different
line-of-sight absorption), and (3) scaling with a ratio of two blackbody models
with different effective temperatures (representing different brightness
and colours). This model approach does not depend on a source model assuming
the intrinsic source to be the same. It is therefore more sensitive to 
incremental changes than modelling approaches where source and absorption are modelled
simultaneously.}
  % results heading (mandatory)
   {
%The day 37.1 observation is much different from the contemporaneous 2006
%observation on day 39.7 resembling much more the earlier-epoch
%observation on day 2006d26.1. These faint-SSS spectra are highly complex
%requiring dedicated discussion, postponed to a future publication.
The 2021d55.6 spectrum can be reproduced remarkably well by
multiplying the (brighter) 2006d39.7 and 2006d54 spectra with the
absorption model, while the 2006d66.9 spectrum requires additional
colour changes to match the 2021.d55.6 spectrum.
The 2006d39.7 spectrum much more closely resembles the 2021d55.6 spectrum in
shape and structure than the same-epoch 2006d54 spectrum:
The spectra on days 2006d39.7 and 2021d55.6 are richer in absorption lines
with a deeper O\,{\sc i} absorption edge,
and blueshifts are higher ($\sim 1200$\,km\,s$^{-1}$) than on day
2006d54 ($\sim 700$\,km\,s$^{-1}$).
In the SSS light curves on days 2006d39.7, 2006d54, and 2021d55.6,
brightness and hardness variations are correlated, indicating variations
of the O\,{\sc i} column density. Only on day 2006d39.7, a
1000s lag is observed.
%The corresponding variations in absorption could be the cause of the
%observed brightness variations, or they could be a consequence of
%changing the degree of ionisation of the absorbing material.
The 35s period was detected on day 2021d55.6 with lower significance
compared to 2006d54.
   }
  % conclusions heading (optional), leave it empty if necessary 
   {
%The fainter SSS emission on day 2021d55.6 compared to the
%earlier-epoch 2006d39.7 and the same-epoch 2006d54 spectra can
%be explained by a higher column density of hot and cold material
%alone.
We conclude that the central radiation source is the same, while
absorption is the principal reason for observing lower soft-X-ray emission in 2021 than in 2006. This is consistent with
a similar 2006 and 2021 [Fe\,{\sc x}] line-flux evolution.
We explain the reduction in line blueshift, depth in O\,{\sc i} edge,
and number of absorption
lines from day 2006d39.7 to 2006d54 by deceleration and heating of the
ejecta within the stellar wind of the companion.
In 2021, less such deceleration and heating was observed, which we
interpret as due to viewing at different angles through an inhomogeneous
density distribution of the stellar wind, allowing free expansion
in some directions (probed in 2021) and a higher degree of deceleration
in others (probed in 2006). The higher absorption in 2021 can then be
explained by the lower-temperature absorbing plasma being more opaque
to soft X-rays.
%the density
%distribution of the stellar wind of the companion compared to the
%earlier outburst which would be consistent with radio observations
%showing different morphology and light curve evolution between 2006 and 2021.
Our approach of scaling observations against observations is free of
ambiguities from imperfect source models and can be applied
to other grating spectra with complex continuum sources.
    }
   \keywords{(stars:) novae, cataclysmic variables - stars: winds, outflows - X-rays: binaries - stars: individual (RS\,Oph)
}

   \maketitle

%
%-------------------------------------------------------------------

\section{Introduction}

During a typical nova outburst, hydrogen is fused to helium on the surface of a white dwarf star. The burning material is obtained from a companion star via accretion, and a variety of companion stars can act as hydrogen donors. Over a period of continuous accretion, hydrogen accumulates on the surface of the white dwarf until ignition conditions are reached leading to thermonuclear runaway reactions. While explosive nuclear burning is highly energetic, white dwarfs are sufficiently sturdy to survive a nova outburst. As opposed to supernovae, novae are thus cataclysmic rather than catastrophic events, and the pre-outburst configuration is essentially preserved. After a nova outburst, accretion continues, eventually leading to another outburst once the ignition conditions are reached again. The recurrence timescale is highly sensitive to the mass of the white dwarf, which determines the ignition conditions and how quickly they can be reached. This means that the interoutburst time for a given white dwarf mass depends to the highest degree on the accretion rate, which is not necessarily constant, and interoutburst times can therefore vary.\\

While all novae are recurrent, the class of recurrent novae is defined as those for which at least two outbursts have been observed. RS\,Oph is one of the most famous recurrent novae, having been observed in outburst with roughly 10-20 years recurrence time since 1898, and the seventh\footnote{A further
possible eruption took place in 1945 \citep{adamakis2011} and was greatly
obscured by the seasonal gap}
confirmed outburst was reported by the AAVSO (American Association of Variable Star Observers) network with Alert Notice 752 on 2021 August 9. The optical peak magnitude of 4.5 was reached around 2021-08-09.5417 UT, which is used as the reference time $t_{\rm ref}$ for this work.
\cite{munari2022} performed weighted spline fitting to the whole set of AAVSO data
resulting in 2021-08-09.58 UT ($\pm 0.05$ days), that is, 0.04 days (=55 minutes)
later than our value of $t_{\rm ref}$ but within the error uncertainty.\\

Nuclear reactions occur in the form of hydrogen burning via the CNO cycle and the energy produced is initially dissipated as high-energy radiation, powering a radiation-driven wind. During the early fireball phase, super-soft-source (SSS) emission is expected to be observable for a short time and was recently observed in serendipitous eROSITA observations \citep{ole}.  Shortly after the initial detonation, the white dwarf will be surrounded by an optically thick ejecta envelope that is initially opaque to high-energy radiation. At the pseudo-photosphere of this envelope, the energy escapes and becomes observable to us in lower-energy bands, depending on the extent of the optically thick part of the ejecta. As the expansion continues and the mass-ejection rate slows down, the pseudo-photospheric radius decreases with time, and the peak of the spectral energy distribution (SED) moves into the UV and soft X-ray bands. Initially, a nova is brightest in the optical bands, later moving into the UV bands, and is ultimately brightest in the soft X-rays, when an atmospheric spectrum of several $10^5$\,K is emitted and observed as an SSS X-ray spectrum. For more details, we refer
to \cite{bodeevansbook}.\\

The intense SSS radiation affects the surrounding medium in ways that facilitate indirect
observational evidence for the presence of SSS emission. \cite{munari2022}
present the evolution of coronal iron lines ([Fe\,{\sc x}],
[Fe\,{\sc xi}], [Fe\,{\sc xiv}]) and of O\,{\sc vi} 1032\,\AA, 1036\,\AA, which are
Raman-scattered to 6825\,\AA, 7088\,\AA\ by neutral hydrogen in the ground state
(see their Fig. 5, and \citealt{schmid89} for identification of the Raman line).
All of these lines rose substantially after day $\stackrel{>}{_\sim}20$ after maximum
optical brightness of the 2021 outburst of RS\,Oph where for lines in higher ionisation stages,
this rise was observed later.
For RS Oph in 2021, the coronal lines remained stable through
days 30-86, as if the number of ionising photons fed to the expanding medium
remained stable until the nuclear burning was switched off.
All coronal lines further decline in parallel with the declining continuum,
as if governed only by dilution in the ejecta under stable illumination by the
central source.
Figure~14 of \cite{page2022} shows a comparison between the evolution
of the [Fe\,{\sc x}] emission line flux (6375\,\AA) and the \swift/XRT soft-X-ray
count rate, demonstrating the temporal correlation of SSS emission and coronal emission
line intensity.\\

Until the start of the SSS phase, the pseudo-photosphere is X-ray dark, but other processes less directly related to the white dwarf can produce X-rays. The companion in RS Oph is a red giant and so produces a dense stellar wind.
The nova ejecta interact with this wind to give rise to a forward shock running into this wind with a reverse shock running into the ejecta. Some of the kinetic energy of the ejecta is thus converted to thermal energy and particle acceleration leading in turn to line and continuum emission across the electromagnetic spectrum, including the X-ray \citep[e.g.][]{bodekahn85,obrien94,vaytet07}.
The sweeping up of the wind of the companion red giant in symbiotic systems such as RS\,Oph, plus the emission of some of the initial energy imparted by the ejecta, lead to deceleration and hence lower-than-usual expansion velocities compared to novae occurring in non-symbiotic systems later in their evolution. This was seen in the 2006 outburst of RS\,Oph by \cite{nessrsoph} for example.\\

Radio interferometric imaging of the 2006 and 2021 outbursts reveals the
structure of the ejecta. A more accurate {\it Gaia} astrometric position allowed \cite{munari_radio}  to obtain a more concise model, which indicates two lobes east and west of the
white dwarf, expanding perpendicular to the orbital plane. These lobes evolved similarly
in 2006 and 2021. Figure~2 in \cite{munari2022} indicates that the flux ratio between
blue- and redshifted peaks reversed between 2006 and 2021, which can be explained
by the orbital phases differing by nearly 180 degrees. Possibly, circular symmetry does not
hold along the inner wall of the density enhancement in the orbital plane, and this may
influence how SSS radiation reached us in 2006 and 2021.\\

During the sixth recorded outburst in 2006, the evolution of RS\,Oph in X-rays was intensively followed with \swift\ \citep{bode06,osborne11}, guiding deeper high-resolution spectroscopy observations with \chandra\ and \xmm\
\citep{drake_rsoph,nessrsoph,rsophshock,nelson07}. Until day $\sim 25$ after optical peak, only the shock emission was observed with,  after
an initial peak, ultimately a slowly declining brightness, while the distribution of electron temperatures gradually shifted, indicating a cooling plasma. Around day 26, the SSS was first observed with a highly variable \swift\ count rate, which was a great surprise at the time. High-amplitude variations were thought to be caused by occultations by clumps within inhomogeneous outer ejecta, although this has been the subject of great debate, and previous studies have proven inconclusive \citep{osborne11}. The SSS high-resolution grating spectra contained deep absorption lines originating from highly ionised elements that were blueshifted by $\sim 800-1000$\,km\,s$^{-1}$, which is slow compared to other novae. The slower expansion velocity could be explained by substantial amounts of kinetic energy having been dissipated within the stellar wind of the companion star.\\

The seventh outburst of RS\,Oph met us with essentially the same X-ray observing capabilities in space, giving us the unique opportunity to study two outbursts with the same instrumentation. With our observing strategy, we aimed to identify similarities and differences but also to follow up questions that remained from the 2006 observation campaigns. The \swift\ observing campaign is described by \cite{page2022}, while the \chandra\ and \xmm\ observations during the shock phase are described by \cite{orio2022}. The expectations motivating the observing strategy are described in \S\ref{sect:obsstrat} where we also describe instrumental issues and how we solved them (with more details in Appendix \ref{appendix}). In \S\ref{sect:analysis} we describe the methods with which we analyse our observations,
combined with the results obtained from the data studying variability (\S\ref{sect:variability}), spectra (\S\ref{sect:analysis:spec}, and short-term spectral variability within the observation (\S\ref{sect:specvar}).
Our results are discussed in \S\ref{sect:disc} and summarised with conclusions in \S\ref{sect:concl}.

\section{Observations}
\label{sect:obs}

\begin{figure*}[!ht]
\resizebox{\hsize}{!}{\includegraphics{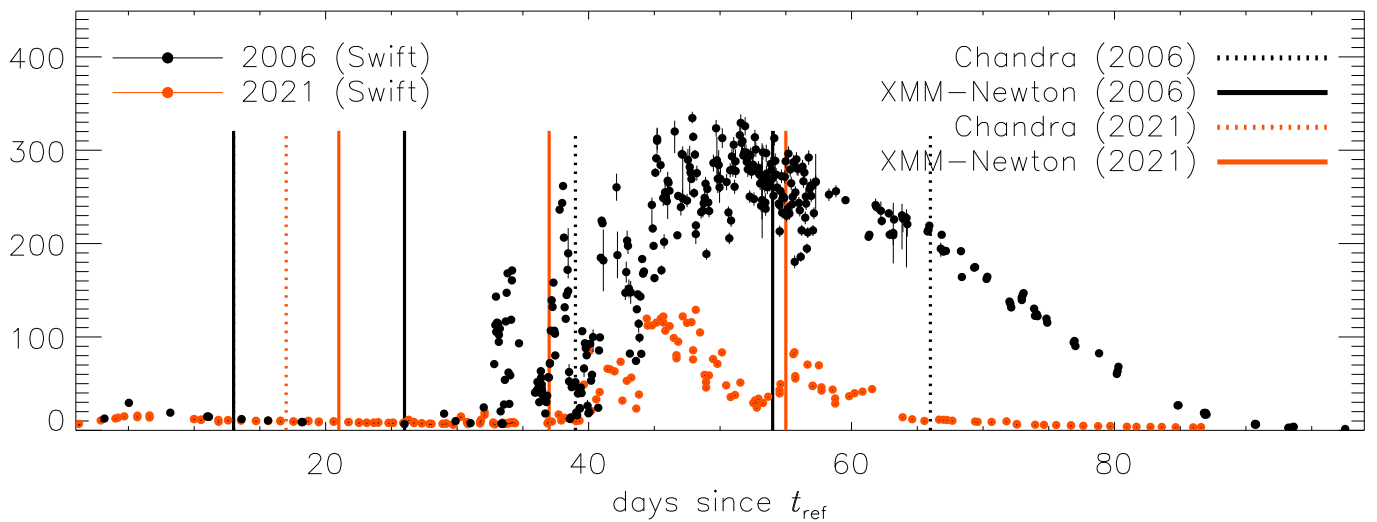}}
        \caption{\label{fig:mmlc}Comparison of long-term \swift\ X-ray light curves obtained during the 2006 (black) and 2021 (orange) outbursts \citep{page2022}. The similar behaviour during the early shock phase until $\sim 30$ days after $t_{\rm ref}$ can be seen while the remarkable differences during the SSS- and decline phases after day $\sim 30$ are evident.
Vertical lines mark the times of deeper \chandra\ (dotted) and \xmm\ (solid)
observations for the outbursts in 2006 (orange) and 2021 (black) as marked in
the upper right legend. In this work, the two \xmm\ observations taken on days 37.1 and 55.6 are discussed.
}
\end{figure*}

\begin{table*}[ht]
\begin{flushleft}
\renewcommand{\arraystretch}{1.1}
\caption{\label{tab:obs}Journal of exposures taken with \xmm\ and \chandra\ (last rows) used in this work}
\begin{tabular}{lllllll}
\hline
ObsID & Instr. & Mode & Filter & Start time & Stop time & exposure time\\
\hline
0841930501 & pn & Timing/Burst & Thick & 2021-09-15T15:07:26 & 2021-09-16T20:33:09 & 105.83\,ks\\
day 37.07$^a$ & RGS1 & \multicolumn{2}{l}{RGS Small Window}  & 2021-09-15T14:38:15 & 2021-09-16T20:37:03 & 107.82\,ks\\
 & RGS2 & \multicolumn{2}{l}{RGS Small Window} & 2021-09-15T14:39:04 & 2021-09-16T20:35:48 & 107.72\,ks\\
0841930901 & pn & Timing & Thick & 2021-10-04T04:18:23 & 2021-10-04T19:14:59 & 25.73\,ks\\
day 55.61$^a$ & MOS1 & Small Window & Thick & 2021-10-04T03:40:18 & 2021-10-04T19:17:50 & 56ks\\
& MOS2 & Timing & Thick & 2021-10-04T03:40:34 & 2021-10-04T19:13:36 & 55.7ks\\
 & RGS1 & \multicolumn{2}{l}{Spectroscopy}  & 2021-10-04T03:40:05 & 2021-10-04T19:18:34 & 56.21\,ks\\
 & RGS2 & \multicolumn{2}{l}{Spectroscopy} & 2021-10-04T03:40:13 & 2021-10-04T19:18:36 & 56.12\,ks\\
 & OM & IMAGE+FAST & UVW2 & 2021-10-04T03:49:07 & 2021-10-04T05:02:27 & 4400\,s\\
 & OM & IMAGE+FAST & UVW2 & 2021-10-04T05:07:34 & 2021-10-04T06:20:54 & 4400\,s\\
 & OM & IMAGE+FAST & UVW2 & 2021-10-04T06:26:00 & 2021-10-04T07:39:20 & 4400\,s\\
 & OM & IMAGE+FAST & UVW2 & 2021-10-04T07:44:28 & 2021-10-04T08:57:48 & 4400\,s\\
 & OM & IMAGE+FAST & UVW2 & 2021-10-04T09:02:55 & 2021-10-04T10:16:15 & 4400\,s\\
 & OM & IMAGE+FAST & UVW2 & 2021-10-04T10:21:22 & 2021-10-04T11:34:42 & 4400\,s\\
 & OM & IMAGE+FAST & UVW2 & 2021-10-04T11:39:48 & 2021-10-04T12:53:09 & 4400\,s\\
 & OM & IMAGE+FAST & UVW2 & 2021-10-04T12:58:15 & 2021-10-04T14:11:37 & 4400\,s\\
 & OM & IMAGE+FAST & UVW2 & 2021-10-04T14:31:42 & 2021-10-04T15:45:03 & 4400\,s\\
 & OM & IMAGE+FAST & UVW2 & 2021-10-04T15:50:10 & 2021-10-04T17:03:30 & 4400\,s\\
 & OM & IMAGE+FAST & UVW2 & 2021-10-04T17:08:37 & 2021-10-04T18:21:58 & 4400\,s\\
\hline
0410180201  & RGS1 & \multicolumn{2}{l}{Spectroscopy} & 2006-03-10T23:02:51 & 2006-03-11T02:20:57 & 11.78\,ks\\
day 26.12$^b$  & RGS2 & \multicolumn{2}{l}{Spectroscopy} & 2006-03-10T23:02:56 & 2006-03-11T02:20:57 & 11.78\,ks\\
0410180301  & RGS1 & \multicolumn{2}{l}{Spectroscopy} & 2006-04-07T21:04:29 & 2006-04-08T02:20:31 & 13.87\,ks\\
day 54.04$^b$& RGS2 & \multicolumn{2}{l}{Spectroscopy} & 2006-04-07T21:04:34 & 2006-04-08T02:20:31 & 13.87\,ks\\
7296 (day 39.7$^b$) & \chandra & LETGS & & 2006-03-24T12:25:22 & 2006-03-24T15:38:20 & 9.97\,ks\\
7297 (day 66.9$^b$) & \chandra & LETGS & & 2006-04-20T17:23:48 & 2006-04-21T09:09:38 & 6.46\,ks\\
%\hline
\end{tabular}
$^a$Observation start in days after $t_{\rm ref}$=2021-08-09.5417
$^b$Observation start in days after 2006-02-12.84
\renewcommand{\arraystretch}{1}
\end{flushleft}
\end{table*}

The details of all observations with exposures used in this work are listed in Table~\ref{tab:obs}. \xmm\ consists of five X-ray telescopes operating in parallel behind three focusing mirrors: The European Photon Imaging Cameras (EPIC) consist of three detectors, the pn detector collecting all the light from one mirror \citep{epic_pn} and two MOS detectors MOS1 and MOS2 \citep{epic_mos} sharing the light of the respective other two mirrors with two Reflection Grating Spectrometers (RGS1 and RGS2; \citealt{rgs}).
The two dispersive RGS spectrometers
%use the high spatial resolution of the underlying MOS detectors to convert the chip positions of each arrived photon to wavelengths based on the properties of the reflection gratings intervening the light beam. This way, the limitation of the low energy resolution of the pn and MOS detectors can be overcome to 
yield high-resolution spectra capable of resolving emission and absorption lines.
In addition, there is an Optical Monitor (OM; \citealt{om}), a 30cm telescope with a CCD detector, six filters, and two grisms. There are two readout modes, an Imaging mode reading out full or parts of the CCD chip of an entire exposure, and a Timing (=Fast) mode, reading out a small area of the chip in event counting mode, recording the coordinates and arrival time of each photon, allowing the construction of a light curve.\\

To distinguish 2006 from 2021 observations, we adopt the naming terminology of {\em year} followed by d and the {\em day} after the respective optical peak, for example 2021d37.1 for the observation taken on day 37.1 after the 2021 optical peak. In addition to the new observations taken on days 2021d37.1 and 2021:55.6, exposure details of RGS exposures for the two archival observations taken on days 2006d26.1 and 2006d54 and the \chandra\ observations taken on days 2006d39.7 and 2006d66.9 are included. The available information for the 2006 outburst gave us a novel opportunity to schedule deeper X-ray grating observations in a pro-active rather than re-active manner in 2021, that is we did not wait until phase changes (such as the start of the SSS phase) were demonstrated with \swift\ but assumed the \swift\ light curve would evolve the same way as in 2006. This enabled us to avoid the delay between trigger and actual \xmm\ or \chandra\ observations. However, we depended on the reality of our assumption, and the strengths and weaknesses of this approach are described in the following section.

\subsection{Expectations motivating the observing strategy}
\label{sect:obsstrat}
As reported in AAVSO Alert notice 752, the duration of the interoutburst interval in RS Oph varies greatly, but the shape of the outburst in optical light and recovery to quiescence is remarkably similar from outburst to outburst. Our baseline expectation was therefore to also see similar behaviour between the 2006 and 2021 outbursts in X-rays and was supported by seeing the results of the first two weeks of \swift\ X-ray monitoring \citep{page2022}. Figure~\ref{fig:mmlc} shows the 2006 and 2021 \swift\ light curves in black and orange colours, respectively, with the times of \chandra\ and \xmm\ observations in 2006 and 2021 marked with the same colours. During the shock phase in 2021, \chandra\ and \xmm\ observations were obtained on days 2021d17.5 and 2021d21.1, respectively, yielding high-resolution X-ray spectra that perfectly fit into the cooling trend seen in 2006 \citep{orio2022}. During the SSS phase, two \xmm\ observations were obtained on days 2021d37.1 and 2021d55.6; these are described in the present work.\\

Assuming the same SSS evolution as in 2006, we made a scheduling request for day 37.1 (ObsID 0841930501, 100ks) with the goal to probe spectral changes during the early SSS phase that had displayed high-amplitude variations in 2006 \citep{osborne11}.
As the peaks of emission were expected to be extremely bright, we selected the most
conservative instrumental setup with RGS in Small Window mode, both MOS instruments
turned off to use their telemetry allocation for downlinking all RGS data, and the pn in burst mode. Unfortunately, the nova did not behave as expected, yielding a much lower count rate, and rendering the restrictive instrumental setup unnecessary. Nevertheless, important new insights can be gathered from this observation when comparing to the observation on 2006d26.1, which will be described in more detail in an upcoming publication.\\

Under the same proposal, we had requested a shorter follow-up observation (approved 47ks) in order to determine the spectral shape after the early variability phase. We scheduled a 56ks observation under ObsID 0841930901 (see Table~\ref{tab:obs}) starting on day 55.6 in order to compare it with the 2006 observation on day 54, thus obtaining two SSS spectra of the same nova during the same epoch of the evolution. Expecting a higher brightness than on day 37, we initially chose the same restrictive instrumental setup of the RGS Small Window Mode and pn burst mode. However, between days 45 and 50, a gradual decline was observed \citep{page2022}, and we therefore changed the instrumental setup to standard RGS spectroscopy mode for the benefit of adding the MOS cameras. Also, because the pn burst mode is not sensitive below 0.7\,keV, we changed the pn to Timing mode. Unfortunately, the brightness started to increase again just hours before the start of the \xmm\ observation, and the RGS suffered some problems owing to the soft brightness, which we describe in \S\ref{sect:obs:reduct}.

\subsection{Data processing}
\label{sect:obs:reduct}

The uncalibrated observation data (Observation Data File=ODF) were downloaded from the ESA \xmm\ archive and processed with the Science Analysis Software (SAS) version 19.0.0 and the latest version of calibration files. For the data of the OM, the two standard chains {\tt omichain} and {\tt omfchain} were run to produce image and fast-mode data, respectively. Owing to the brightness in the soft-X-ray band, all X-ray exposures suffered some type of anomalies, which we dealt with thanks to the help of the \xmm\ helpdesk as described in this section, with more details in the Appendix.\\

The RGS Small Window mode data in ObsID 0841930501 suffered some issues with
an inconsistency flagged in the SAS terminology as `GTI\&EXPOSU Inconsistency', which cannot be handled by the SOC. A workaround was found to obtain calibrated
spectra and light curves and will be described in an upcoming paper.
Our choice of instrument setup for ObsID 0841930901 was strongly
influenced by these issues, coupled with the declining rather than increasing
\swift\ count rate.
As described above, the source became brighter again just before the start
of the \xmm\ observation, and although the source was nowhere near as bright
on day 2021d55.6 as it was on day 2006d54 (see Fig.~\ref{fig:mmlc}),
we suffered more problems with ObsID 0841930901 than we had in 2006
with ObsID 0410180301.\\

The RGS2 spectrum is not usable owing to excessive pile up in addition to the loss of
an unknown number of events. The \xmm\ helpdesk provided the following
background information:
Since August 2007, the RGS2 CCDs are read via a single node because of an electronics malfunction. RGS2 frame times are therefore twice as long as before this change, now with an accumulation time of 9.6s in normal spectroscopy mode and 2.4s for Small Window mode when reading the eight CCDs. Therefore, the RGS2 is now more vulnerable to excessive pileup.\\

Also, the RGS1 is affected by pileup, but due to the still shorter readout time, not in a way that photons were lost in an uncontrollable way as in the RGS2. As photons are only re-distributed, we can correct the RGS1 data for pile up as described in \S\ref{append:pileup}. We therefore only use the pile-up-corrected RGS1 spectrum.
Unfortunately, this leaves a larger number of gaps in the grating spectrum, because the RGS1 bad pixels cannot be filled with information from the RGS2. We note that the fluxed RGS spectrum in the pipeline products for ObsID 0841930901, {\tt P0841930901RGX000FLUXED1003.FTZ}
is invalid, yielding an artificial flux reduction in the range 24-28\,\AA. As reported by \cite{nessrsoph}, the 2006d54 spectrum was also piled up, but both RGS2 and RGS1 spectra could be corrected with the same approach, which is described in \S\ref{append:pileup}.\\

Another issue is telemetry losses that can lead to aliases in the period analysis.
In \S\ref{append:TM}, we describe our checks, finding
aliases at $\sim 24$s and $\sim 28$s  in the pn light curve, while there is no other alias expected, and therefore 
the feature in the periodogram at 35.47-s period is not related
to the telemetry dropouts and can be associated to the source
(see \S\ref{sect:variability}).

\section{Analysis and results}
\label{sect:analysis}

\begin{figure}[!ht]
\resizebox{\hsize}{!}{\includegraphics{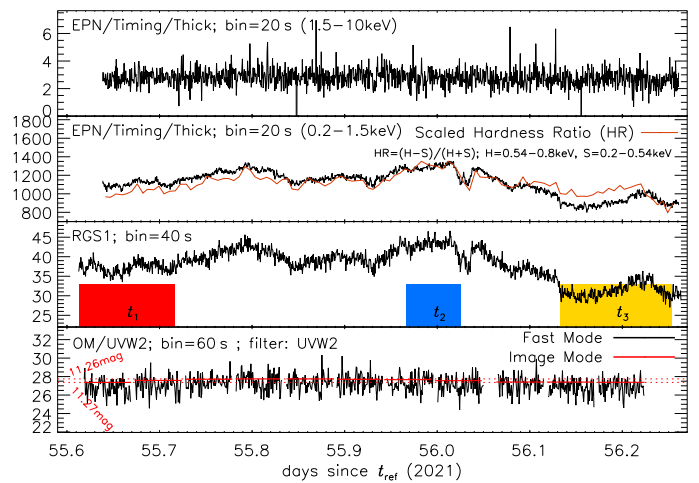}}
	\caption[multilc]{\label{fig:xmmlc}\xmm\ X-ray and UV light curves (units counts per second, cps) extracted from the pn, RGS1, and OM instruments taken on day 2021d55.6 (see x-axis labels). The photon arrival times were converted to days after $t_{\rm ref}$=2021-08-09.5417. The pn light curve is broken into a hard (top) and a soft (second panel) band split at 1.5\,keV (see legends for individual energy bands), probing shock and SSS emission, respectively. The soft band is split into two bands, $H$ and $S$, with the split at 0.54\,keV, and the orange curve in the second panel represents the scaled hardness ratio ($HR$) constructed from countrates recorded in the energy bands $H$ and $S$ as given in the legend. The shaded coloured area in the third panel marks light-curve segments discussed separately in \S\ref{sect:specvar}. In the bottom panel, the OM UVM2 filter ($\sim 2070-2170$\,\AA\ range) light curves are shown from the Image mode (red horizontal bars for each exposure) and Fast mode (black) data. Count rates are plotted, and corresponding magnitude values are included on the left, which correspond to the lowest and highest Image mode count rates (marked with the dotted horizontal red lines).
}
\end{figure}

\subsection{Variability studies}
\label{sect:variability}

Figure~\ref{fig:xmmlc} shows the X-ray and UV light curves extracted
from the pn, RGS1, and OM exposures as labelled in the respective panel legends.
The hard-X-ray light curve (top panel) whose emission originates in the shocks
is flat, demonstrating that all variability originates in the SSS component.
Between days 55.6 and 56.3, random variability of a factor of $< 2$ can be seen
consistently in the soft pn and RGS light curves. Towards the end of the
observation, the RGS1 and pn count rates are $\sim 20$\% lower,
starting with a continuous decline around day 56, slowly rising again around
day 56.2, and dropping slightly back again.
The scaled hardness ratio curve is overplotted in
the second panel (see discussion in \S\ref{sect:specvar}).
The RGS2 light curve on day 55.6 (not shown) is severely compromised by
telemetry losses
and pile up, yielding count rates of between 8 and 16 counts per second (cps).
This is much lower than the count rate in the RGS1 light curve of namely 30-40cps,
and the trends seen in the RGS1 and pn light curves in Fig.~\ref{fig:xmmlc}
are not seen in the RGS2 light curve.\\

\begin{figure}[!ht]
        \resizebox{\hsize}{!}{\rotatebox{270}{\includegraphics{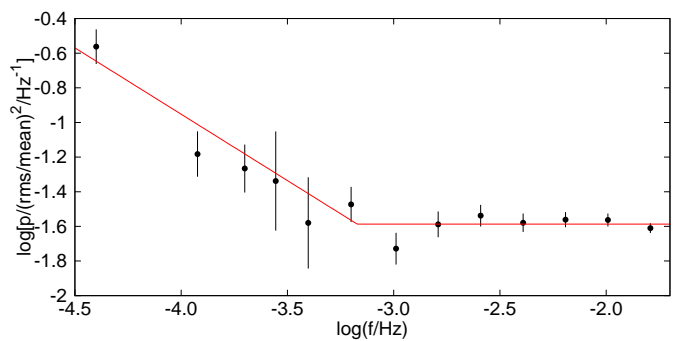}}}
        \caption{\label{fig:power_pn}PDS calculated from the pn data (20s binning) where the red solid line is a broken power-law fit. The circles are average values within the frequency bin with error bars as error of the mean. The frequency range corresponds to a period range of 100-32000 seconds.
        }
\end{figure}

Figure~\ref{fig:power_pn} shows the power density spectrum (PDS) describing random variability in the pn light curve of timescales longer than 100s. The light curve was split into four\footnote{We also tried splitting into  two or three subsamples to expand the frequency range but the PDSs are more noisy. Larger binning is needed, which decreases the frequency range. Otherwise the results are robust.} subsamples, a log-log periodogram\footnote{The Lomb-Scargle algorithm in the Python package {\tt Astropy}
(\citealt{astropy_collaboration2013,astropy_collaboration2018,astropy_collaboration2022})
was used. Periodograms were rms normalised (\citet{miyamoto1991}).} was calculated from each subsample, and the resulting periodogram points were binned with a bin width of 0.2 in dex. All periodogram points within a bin were averaged and the error of the mean was calculated. It can be seen that the low-frequency part of the PDS is dominated by red noise\footnote{described by a broken power-law fit, with a constant level above the break frequency.} up to frequency log($f$/Hz) =
-3.17, which corresponds to a timescale of 0.017 days (25 minutes).
All variability longer than this timescale belongs to this red noise.
No break or other characteristic frequencies are detected.\\

\begin{figure}[!ht]
        \resizebox{\hsize}{!}{\rotatebox{270}{\includegraphics{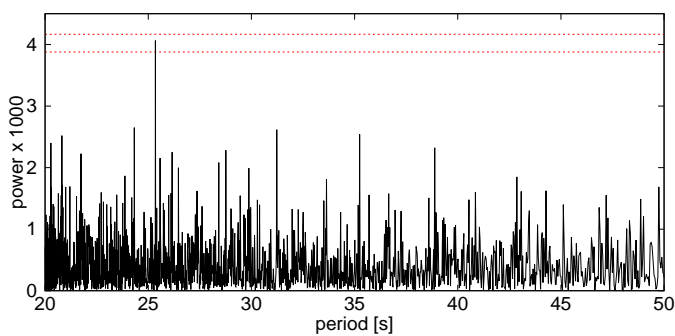}}}
\caption{\label{fig:power_om}Periodogram calculated from the OM data with 70\% and 50\% confidence levels as upper and lower horizontal dashed lines, respectively. The PDS standard normalisation of the Lomb-Scargle algorithm was used for the confidence level calculation.
        }
\end{figure}

The UVW2 Image mode magnitudes are shown in the bottom panel of
Fig.~\ref{fig:xmmlc} with 11 red horizontal bars whose lengths correspond
to the respective integration times of 4400s each. These magnitudes vary in a
narrow range of 11.27-11.26 mag (red dotted horizontal lines),
and within this range, no systematic decline or
increase can be identified. We further found no
periodic variability in the OM Fast mode light curve (black) as can be
seen from the power spectrum shown in Fig.~\ref{fig:power_om}. The
highest peak is below 70\% confidence and is therefore probably a random event.
In particular, there is no signal at the 35s period that was seen
in X-ray light curves.\\

\begin{figure*}[!ht]
\resizebox{\hsize}{!}{\includegraphics{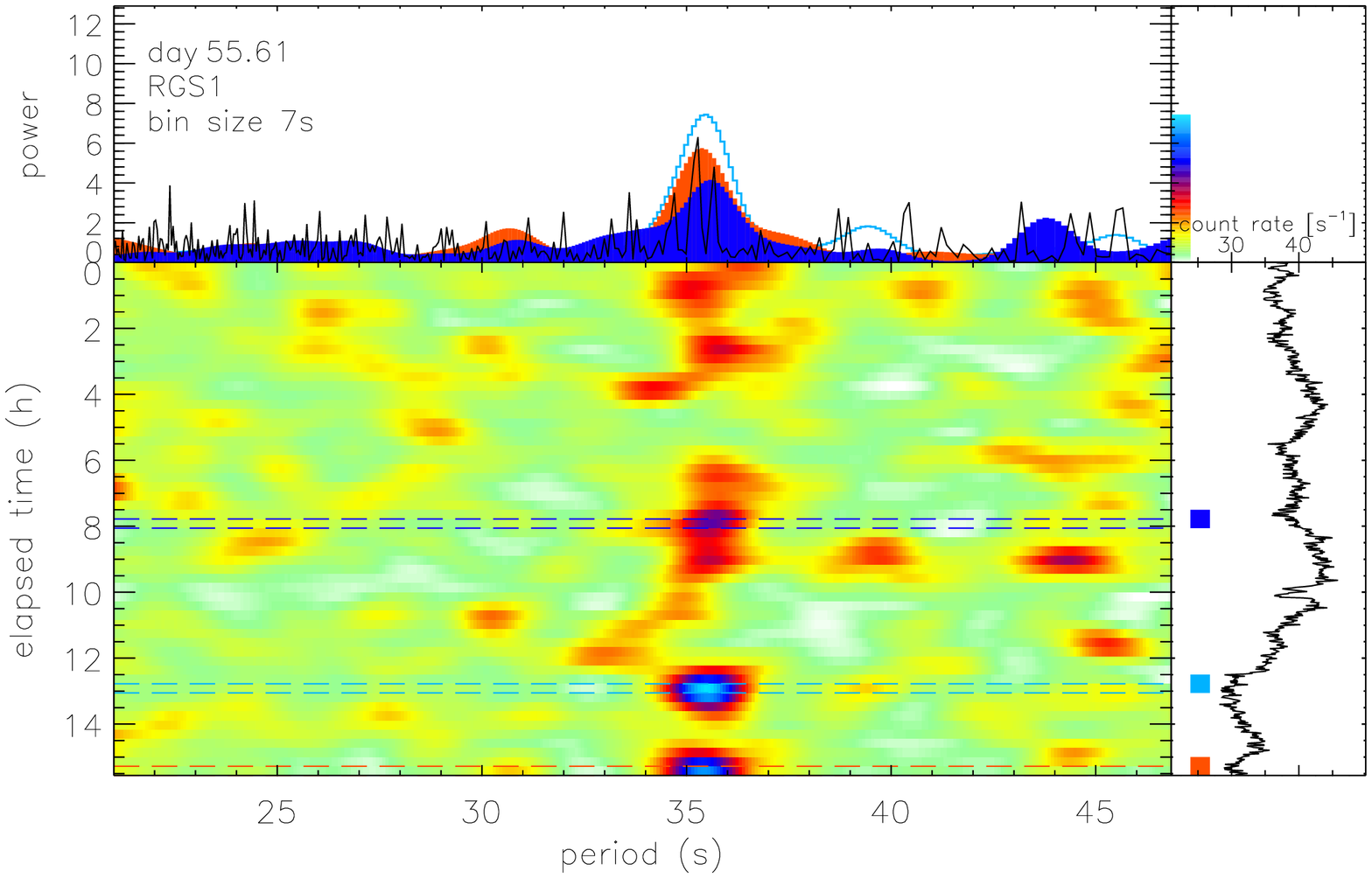}\includegraphics{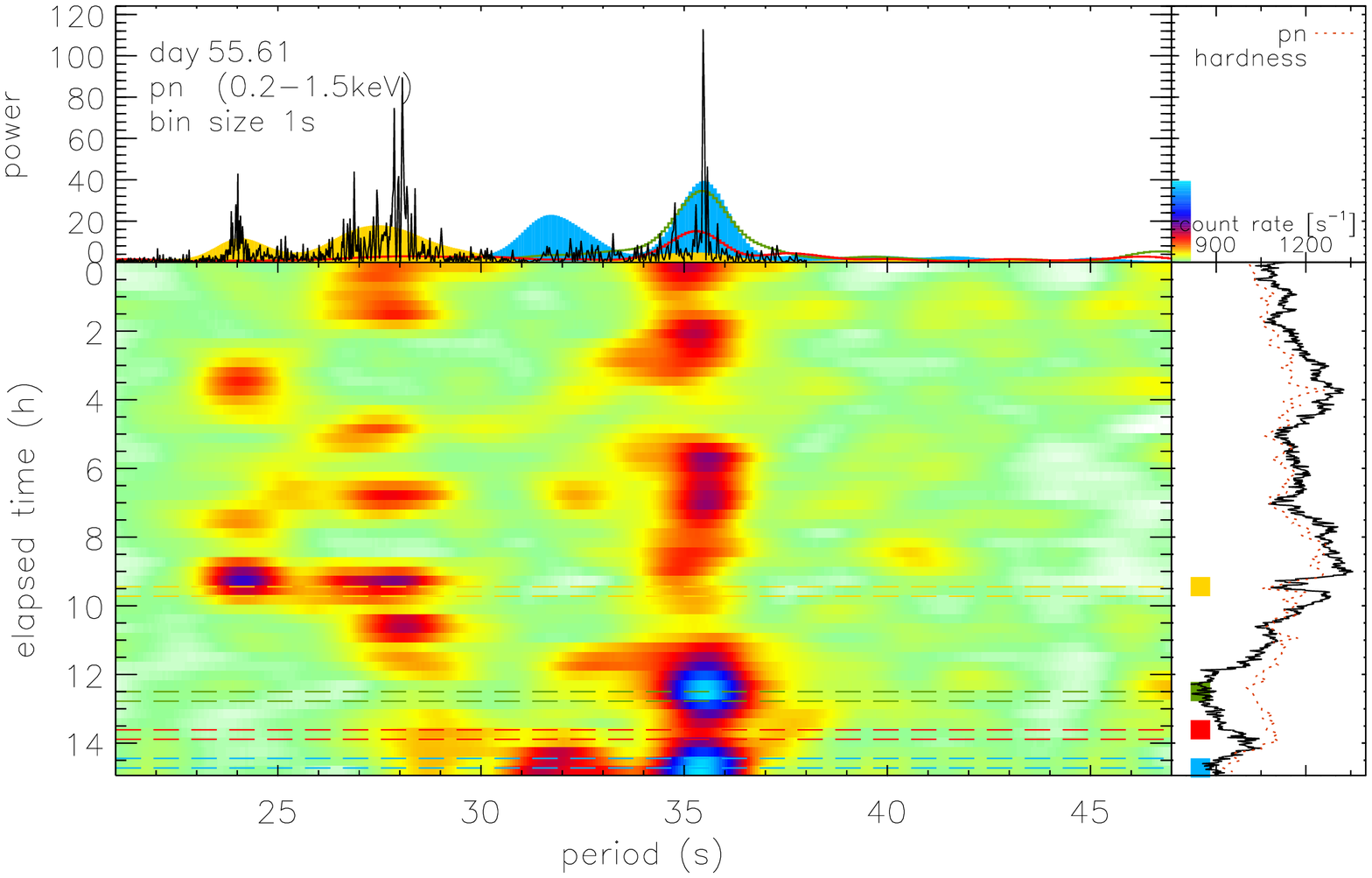}}
\caption{\label{fig:lmap}\xmm\ RGS (left) and pn (right, soft 0.2-1.5keV range)
periodogram time maps illustrating the evolution of periods focusing on the 35s period detected during the 2006 outburst.
Each graph consists of four panels: in the main (bottom left) panel, time
is running down and the tested periods run from left to right, while the panel above shows
representative Lomb-Scargle periodograms on the same shared horizontal axis.
The bottom-right panel shows the light curve along the same vertical time
axis. The colour bar in the top right allows conversion of colours in the
main panel to the units of power used for the periodograms. The vertical
time axis is in units of elapsed hours since the start of the RGS1 exposure (left
panel), and the pn exposure (right panel) which started 2250s (0.6 hours) later than
the RGS1 exposure owing to the higher pn overhead. In the respective top panels, the black
curves show the
periodograms from the respective entire light curves and the colour-shaded
curves are periodograms from the selected 2000s time intervals marked with
horizontal lines of the same colour as the dashed lines in the panels below.
The time bin sizes
of the light curves are included in the respective top-left panels. The
same time maps were produced with different bin sizes, yielding similar results.
}
\end{figure*}

During the 2006 outburst, transient 35s periodic oscillations were
detected with both \swift\ \citep{atel770, osborne11} and \xmm\
\citep{nelson07,nessrsoph,nessqpo}. This period was only present in
the plasma that emits the SSS spectrum \citep{nessrsoph}. This period was again found with
\swift\ during the 2021 outburst \citep{page2022}, indicating that it
is related to system parameters.
We searched for this period in the soft pn and RGS1 light curves from day 55.6. 
As it was transient, we studied the time evolution following the methods
described in \cite{nessqpo}, splitting each light curve into 54 (soft pn)
and 55 (RGS1) adjacent overlapping subintervals of 2000s duration with 50\%
overlap. For each of these subintervals, we calculated separate Lomb-Scargle power
spectra with the method of \cite{scargle1982}.
In Fig.~\ref{fig:lmap}, period time maps are shown where a clear transient
period can be seen around 35s in both
RGS1 and pn light curves. We repeated the same study with different
bin sizes, all yielding consistent results.
We also detected 35s oscillations in the RGS2 light curve, even
though it is seriously compromised by pile up and telemetry losses.
The trends in the RGS2 light curve are different from those in the pn
and RGS1 light curves, which we attribute to the higher levels of pile
up and telemetry loss due to the longer read-out time.\\

The period is variable in strength but this variability is not obviously correlated
with the count rate. Towards
the end of the observation, the strongest signal is detected,
when the count rate was $\sim 20$\% lower. During this last part
of the observation, a small rise in count rate is seen that seems to
correlate with a much lower power of the period. It is possible that during
the last part of the observation, the variability behaviour changed,
yielding an anti-correlation between period power and count rate. We
remind the reader that the correlation between hardness and
count rate was less clear towards the end of the observation
(see Fig.~\ref{fig:xmmlc}).\\

The total pn light curve
shows two strong signals around 24s and 28s not seen in the RGS1
data. This is caused by  the pn instrument
 reaching telemetry saturation (see \S\ref{sect:obs:reduct} and Fig.\ref{fig:dbin}).

\subsection{Spectral studies}
\label{sect:analysis:spec}

\begin{figure}[!ht]
\resizebox{\hsize}{!}{\includegraphics{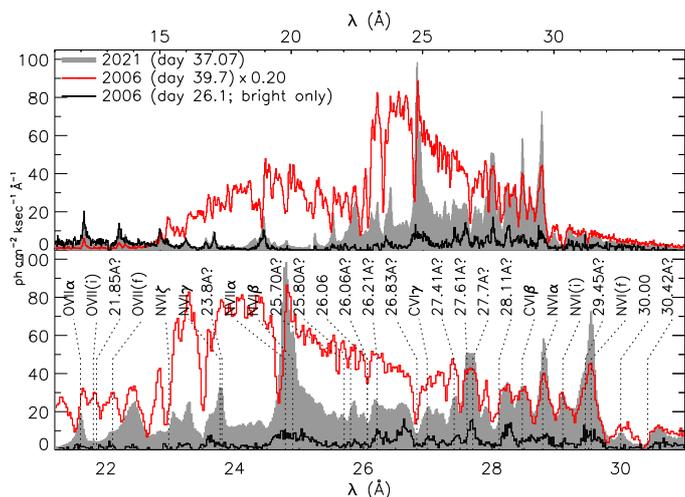}}
\caption{\label{fig:xmmspec37}\xmm\ RGS spectra in photon flux units per
10$^3$\,s (=ksec). The 2021d37.1
spectrum (grey shading) is compared to the same-epoch 2006d39.7 spectrum
(down-scaled by factor 5; red) and the bright-only 2006 spectrum on
day 26.1 (black). The top panel shows a larger wavelength range
while the bottom panel zooms into the variable soft excess with several
unknown narrow emission and absorption features, where only the corresponding
wavelength is indicated.
}
\end{figure}

The two spectra taken on days 2021d37.1 and 2021d55.6 probe different
phases of the evolution. In Fig.~\ref{fig:xmmspec37} we compare the
RGS spectrum from day 2021d37.1 (grey shading) with the same-epoch
2006d39.7 spectrum down-scaled by a factor five (red), and the earlier
2006d26.1 spectrum extracted from the times after a sudden brightness increase
(black). The top panel shows a larger wavelength range demonstrating
that the same-epoch 2006d39.7 spectrum is very different,
while more, qualitative similarities are seen with the 2006d26.1 spectrum.
The shock emission was the same, while the soft
component with many unidentified narrow emission features was
much brighter on day 2021d37.1. The bottom panel zooms into the soft
range, demonstrating that qualitatively, many of the unknown features
on day 2006d26.1 were also present on day 2021d37.1.
Many of the unknown features are also present in the 2006d39.7
spectrum. These spectra demonstrate that a dedicated
publication is needed to discuss these complex spectra in sufficient
depth. The 2021d37.1 data are therefore not discussed further in this work
with reference to an upcoming publication.\\

In this work, we concentrate on the day 2021d55.6 spectrum,
focusing on identifying reasons for the lower brightness of the SSS
emission compared to 2006d39.7, 2006d54, and 2006d66.9. We
test three hypotheses and any combination of them: simply scaling with a
constant factor (e.g. caused by occultations or eclipses or by a
lower nuclear burning rate), a change
of effective temperature (or colour change), and more absorption.\\

We present a novel approach, applying relatively simple models
to observations rather than source models to scale observations
against observations. With this approach, all the known and
unknown details that are present in both observed spectra are mapped
against each other without needing a full description of the processes
leading to all observed details.\\

The standard approach is to fit spectral models to the data comparing
model with data. Physical quantities such as the effective temperatures or
mass of the underlying white dwarf can be determined this way, which in
turn are useful for constraining evolutionary models needed to obtain a
full understanding of the nova phenomenon.
However, robust physical conclusions are only possible if the
spectral models reproduce the data in full detail, which has so far not
been achieved with the SSS grating spectra. The calculation of robust parameter
uncertainties is needed for the interpretation of best-fit parameters,
and this is only possible if a formally acceptable fit (not just `best' fit)
is found, indicated by finding a value of reduced $\chi^2$ near unity.
The most promising results have been achieved with the {\tt SPEX} model first
attempted by \cite{pinto12} and
more recently by \cite{ness_v3890sgr}. While the agreement of model with
data is the best ever achieved for an SSS spectrum, formal statistical
values of reduced $\chi^2$ still range around six, and the models make
some simplifying assumptions, only allowing insights into the properties
of the absorbing layers without allowing determination of the mass of
the white dwarf. We do not follow this approach in this work. It is desirable
to conduct a systematic, homogeneous analysis of a sample of SSS spectra
with the same analysis strategy, the same version of SPEX, and so on to enable comparison
of different systems.\\

Our novel approach is only possible with the grating spectra because of
their low degree of photon energy redistribution compared to the \xmm\ EPIC
cameras or the \swift/XRT. The instrumental response characterises photon
redistribution and calibrates physical flux to instrumental counts.
The conversion from a fluxed model spectrum to a predicted count spectrum
is done via matrix multiplication of the model with a response
matrix that is the result of the instrument calibration. This matrix contains the
largest conversion values on its diagonal, and the lower the spectral
resolution (and thus the larger the amount of photon redistribution),
the more off-diagonal positions are populated with non-zero values.
Near-diagonal response
matrices can be inverted, and high-resolution spectra such as optical
spectra or X-ray grating spectra can be converted to physical flux spectra
by multiplying an observed count spectrum with the inverse response matrix.
For highly populated response matrices of low-resolution instruments, such as the
\xmm\ EPIC cameras or the \swift/XRT, the inversion solutions are far less
unique, and models have to be converted
to instrumental units by smearing out sharp features to the low instrumental
resolution.\\

Our novel approach is to break the separation of models and data and multiply
a multiplicative scaling model (without units) with an observation
in physical flux units (converted from the instrumental counts) and compare
this with another observed spectrum in physical flux units.\\

We performed parameter optimisation, yielding a minimum
value of $\chi^2=\sum \frac{(model-data)^2}{err^2}$, where $model$
is a 2006 observed spectrum multiplied by the down-scaling model: either
a wavelength-independent (i.e. grey) constant factor, the ratio of two blackbodies
(BB-ratio, described in \S\ref{sect:analysis:bbscal}), or an absorption model
(described in \S\ref{sect:analysis:absscal}). Here, $data$ is the
2021d55.6 spectrum; as it has hot pixel gaps at different wavelengths
from the 2006 spectra, we interpolated each 2006 spectrum to the wavelength
grid of day 2021d55.6. The measurement errors $err$ were combined as
the summed squared errors from day 2021d55.6 and the respective 2006 errors.\\

\subsubsection{Scaling by a constant factor}
\label{sect:analysis:scal}

In order to account for wavelength-independent effects, such as eclipses or
other grey occultations, we allow the application of a constant factor. 
This is the simplest scaling process. This factor can also be interpreted
as a luminosity change; for example, if there is a temperature change
(\S\ref{sect:analysis:bbscal}) combined with a change of radius, the scaling factor would correspond to a ratio in radius
$R_2/R_1=\sqrt{scal}$. As the approaches of temperature change
(\S\ref{sect:analysis:bbscal}) and absorption (\S\ref{sect:analysis:absscal})
combine brightness changes with spectral changes, we keep grey scaling
fixed at unity during initial experiments with those models while
testing the additional scaling only to judge the quality of the other
approaches.

\subsubsection{Blackbody scaling}
\label{sect:analysis:bbscal}

\begin{figure}[!ht]
\resizebox{\hsize}{!}{\includegraphics{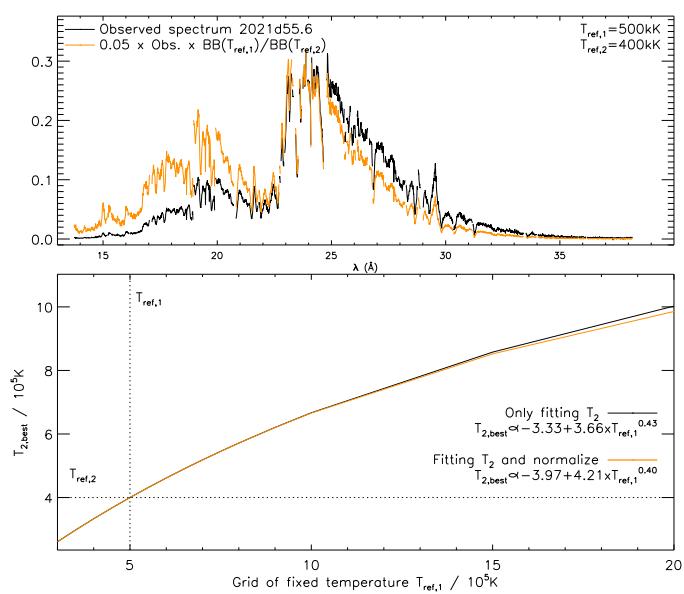}}
        \caption{\label{fig:bbtest}Illustration of the scaling method with the
        ratio of two different blackbody curves and that many pairs of
        $T_{\rm ref,1}$ and $T_{\rm ref,2}$ can produce the same blackbody
        ratio curve. In the top
        panel, the RGS1 spectrum of RS\,Oph from 2021d55.6 (black) is multiplied
        with the ratio of two blackbody models corresponding to
        $T_{\rm ref,1}=5\times 10^5$\,K (=500kK) and $T_{\rm ref,2}=4\times 10^5$\,K
        (=400kK), respectively and normalised (see legend). The resulting spectrum
        (orange) is a factor 20 brighter but is also harder. The bottom panel shows the
        results of fitting only $T_{\rm ref,2}$ with (black) and without (orange)
        additional normalisation while $T_{\rm ref,1}$ is fixed at
        given grid values (x-axis) to reproduce the 500kK/400kK ratio curve.
        The bottom right legend provides empirical analytic relationships
        between fixed grid value and $T_{\rm 2,best}$ values reproducing the
        same ratio curve. The same relationships are found for any other
        pairs of reference temperatures $T_{\rm ref,1}$ and $T_{\rm ref,2}$;
        see \S\ref{sect:analysis:bbscal} for details.
}
\end{figure}

Given the Stefan-Boltzmann law of $L\propto T^4\times R^2$, a temperature
change at the same emitting radius not only shifts the spectrum
to the red or blue but also leads to a substantial change in
brightness. Without the need
for a constant scaling factor (\S\ref{sect:analysis:scal}), small temperature changes
can therefore already lead to large changes in flux. If temperature changes were the
cause of the lower fluxes in 2021 compared to 2006, this would be detectable
as different spectral
shapes. This is illustrated in the top panel of Fig.~\ref{fig:bbtest} where
the observed RGS1 spectrum of RS\,Oph on day 2021d55.6 (black) is
compared with the same spectrum multiplied by the ratio of two blackbody
curves corresponding to $T_{\rm ref,1}=5\times 10^5$\,K and
$T_{\rm ref,2}=4\times 10^5$\,K, respectively (labelled in the legend in
units kK=$10^3$\,K). The resulting spectrum is 20 times brighter at the peak,
and it is therefore renormalised by a factor 0.05 to show that it is harder than the
original spectrum. Multiplication of such ratio curves to the
2006 spectra can therefore probe whether the 2021d55.6 spectrum
might be the result of lower effective temperature by fitting the two
temperatures to yield a best fit of scaled 2006 spectra to the observed
2021d55.6 spectrum. To understand the behaviour of such fits, we stepped
through a grid of fixed values of $T_{\rm ref,1}$ and fitted $T_{\rm 2,best}$
to obtain the closest match to the $500$kK$/400$kK ratio curve and found
that with any value of $T_{\rm ref,1}$ that yields significant
flux within the RGS spectral range ($T_{\rm eff}\approx (3-25)\times 10^5$\,K)
a value of $T_{\rm 2,best}$ can be found to almost exactly reproduce the
$500$kK$/400$kK ratio curve. While fitting both reference temperatures
(\S\ref{sect:spectra}),
we indeed encountered no convergence, and we only managed to constrain
one temperature value if the other is assumed fixed. The respective
best-fit values of $T_{\rm 2,best}$ are shown in the bottom panel
of Fig.~\ref{fig:bbtest}.
The results are very close when allowing additional renormalisation (orange
curves). The bottom panel shows that with higher
values of $T_{\rm ref,1}$, higher values of $T_{\rm 2,best}$ are also
needed. There is no linear relationship, and the bottom-right legend
provides empirical relationships, and these relation ships are in fact
independent of the original choices of $T_{\rm ref,1}$ and $T_{\rm ref,2}$.\\

The conclusion that we draw from this study can be summarised as follows:
\begin{itemize}
\item Down-scaling the 2006 spectra with the ratio of two
blackbody curves corresponding to different temperatures can
explain the lower fluxes in 2021 if a good match can be found
after iterating $T_{\rm ref,1}$ and $T_{\rm ref,2}$.
\item The values of $T_{\rm ref,1}$ and $T_{\rm ref,2}$
are degenerate, and therefore many pairs of $T_{\rm ref,1}$ and $T_{\rm ref,2}$
can produce the same blackbody ratio curve; one of these two
values has to be fixed to any value within the range of
$\sim (8-25)\times 10^5$\,K.
\item When applying this method to data (\S\ref{sect:spectra}),
the resulting values of temperature only inform us as to whether the
compared spectra are qualitatively
cooler or hotter. The method does not provide any concrete
values of respective effective temperatures, and also differences
or relative changes in temperature cannot be quantified.
\end{itemize}

\subsubsection{Absorption scaling}
\label{sect:analysis:absscal}

\begin{figure}[!ht]
\resizebox{\hsize}{!}{\includegraphics{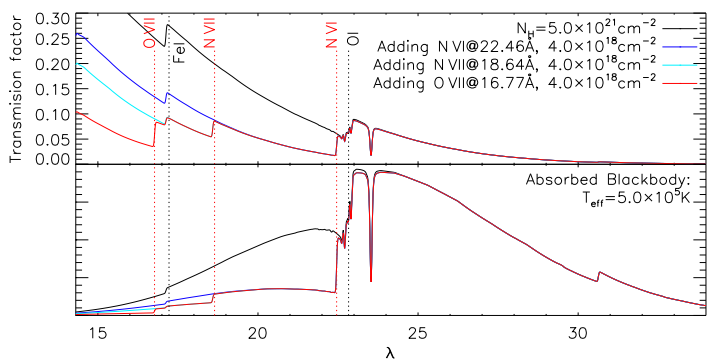}}
        \caption{\label{fig:nhmodel}Illustration of how to down-scale an observed
        source spectrum, assuming higher absorption. In the top panel, transmission
        curves are shown assuming an arbitrary value of neutral hydrogen column
        density $N_{\rm H}$ and cosmic abundances of other elements. Different colours
        represent a pure (neutral) interstellar absorption model (black) and
        modifications when successively adding (hot) absorption edges at the
        respective ionisation
        energies of O\,{\sc vii}, N\,{\sc vii}, and N\,{\sc vi} (see labels); thus,
        the red curve represents the absorption model accounting for all edges. In the
        bottom panel, the effect of absorption is illustrated for the example of
        a blackbody source with an effective temperature of 500,000K. These absorption
        models were designed to be applied to emission models but can also be
        applied to data of sufficiently high spectral resolution;
        see \S\ref{sect:analysis:absscal} for details.
}
\end{figure}

Photoelectric absorption affects photons with energies higher than the ionisation
energies of the intervening material. The ionisation energies increase with the
stage of ionisation, and an ionised plasma is more transparent to X-rays than
neutral material.
To test a higher degree of photoelectric absorption as the cause of the differences,
we multiply the brighter 2006
spectra on days 39.7, 54, and 66.9 with a photoelectric absorption model.
We use the {\tt ionabs} absorption model, which is
part of the PintOfAle package \citep{pintofale} and is illustrated in
Fig.~\ref{fig:nhmodel}. Assuming elemental abundances and an ionisation balance
for each element,  {\tt ionabs}  calculates cross sections from which transmission factors
at each wavelength can be computed that are to be multiplied with a source spectrum.
This model works similarly to the better-known multiplicative model {\tt tbvarabs}
in xspec, which is designed to be applied to additive source models (in flux units) before
folding through an instrumental response and comparison to an observed spectrum in
instrumental units.\\

As the RGS response matrix is sufficiently diagonal, we can afford to
apply the absorption model to the fluxed RGS spectra without the need for
defining (and understanding)
a model of an emission source. As an approximation of the small amount of
photon redistribution, we fold the absorption model with a Gaussian
filter of 0.05\,\AA\ width, which smoothes out the sharp edges and absorption
lines to roughly RGS resolution. This is not possible with CCD spectra as the
much larger photon
redistribution is a statistical process that needs to be modelled.\\

The {\tt ionabs} model calculates photoelectric
absorption cross sections $\sigma_\lambda$ (in units cm$^2$) for a given
photon energy $E={\rm h}\times {\rm c}/\lambda$ for
specified chemical composition (we used here \citealt{agrev89}) and ion fractions.
It uses ground-state photoionisation cross sections computed using the fortran
code of \cite{Verner1996} and is supplemented with high-resolution cross sections
in the vicinity of 1s-2p resonances for O from \cite{Garcia2005}.
%In the current version, fine structure of Oxygen is only
%included for ionisation stages {\sc i-v} while for O\,{\sc vi} and O\,{\sc vii},
%only artificial edges at the ionisation energy are included.\\

From $\sigma_\lambda$, transmission curves
$T_\lambda=e^{-N_{\rm H}\times \sigma_\lambda}$
 as shown in the top panel of Fig.~\ref{fig:nhmodel} can be computed for
an assumed column density $N_{\rm H}$ in units of  cm$^{-2}$. The vector
$T_\lambda$ consists
of factors $\le 1$ for each wavelength value $\lambda$ to be multiplied with
a source spectrum to simulate what would be observed behind an assumed
equivalent hydrogen column of $N_{\rm H}$. The column densities of other
elements are scaled by the assumed abundance relative to hydrogen when
computing $\sigma_\lambda$.\\

The black curve in the top panel of Fig.~\ref{fig:nhmodel} represents
$T_\lambda$, assuming a neutral absorber with ion fraction values of one
for all neutral ionisation stages and zero for higher ionisation stages.
For illustration purposes (to better see the ionisation edges at shorter
wavelengths), the O\,{\sc i} edge was decreased by reducing the oxygen abundance.
In our analysis, we test such modifications equivalent to a change
of the O\,{\sc i} column density. A shallower O\,{\sc i} edge can be interpreted
as a reduced oxygen abundance but may also mean that oxygen is ionised, thus only reducing
the column density of {\em neutral} oxygen. We tested the behaviour of the
transmission curve when the column densities of correspondingly higher ionisation
stages increase, finding that the edge shifts to shorter wavelengths. If this
is not observed, then a reduced O\,{\sc i} edge could be produced by fully
ionising oxygen. Oxygen may also be hidden in other ways; for example, if locked in dust
grains that change the absorbing behaviour to not affect the O\,{\sc i} edge.
We further include absorption by highly ionised material, primarily leading to
edges at 16.77\,\AA\ (O\,{\sc vii}), 18.64\,\AA\ (N\,{\sc vii}), and 22.46\,\AA\
(N\,{\sc vi}), which are marked by red labels in Fig.~\ref{fig:nhmodel}. While
{\tt ionabs} allows neutral and ionised absorption to be included in one
model, we define separate models for each ion, meaning that we obtain the product of
up to four {\tt ionabs} components. One reason for the separation is that the
interstellar matter contains no O\,{\sc vii} and so on, and the ionised material
must be based on a different value of $N_{\rm H}$. It would be more self-consistent
to combine at least the N\,{\sc vi} and N\,{\sc vii} edges into one component.
Fine-tuning the ionisation balance would also require assuming
less-ionised nitrogen, which increases complexity. This would be a different
study from that presented here, where our goal is to identify the role of absorption as an
explanation for the lower fluxes in 2021 compared to 2006. In the bottom panel of Fig.~\ref{fig:nhmodel}, we demonstrate the effects
of the absorption models from the top panel of the same figure on a blackbody source spectrum with
$T_{\rm eff}=5\times 10^5$\,K.
%JOKE
%It does not only resemble the \swift\ light curve (Fig.~\ref{fig:mmlc}) but
%also typical observed SSS spectra.\\

\begin{figure*}[!ht]
        \resizebox{\hsize}{!}{\includegraphics{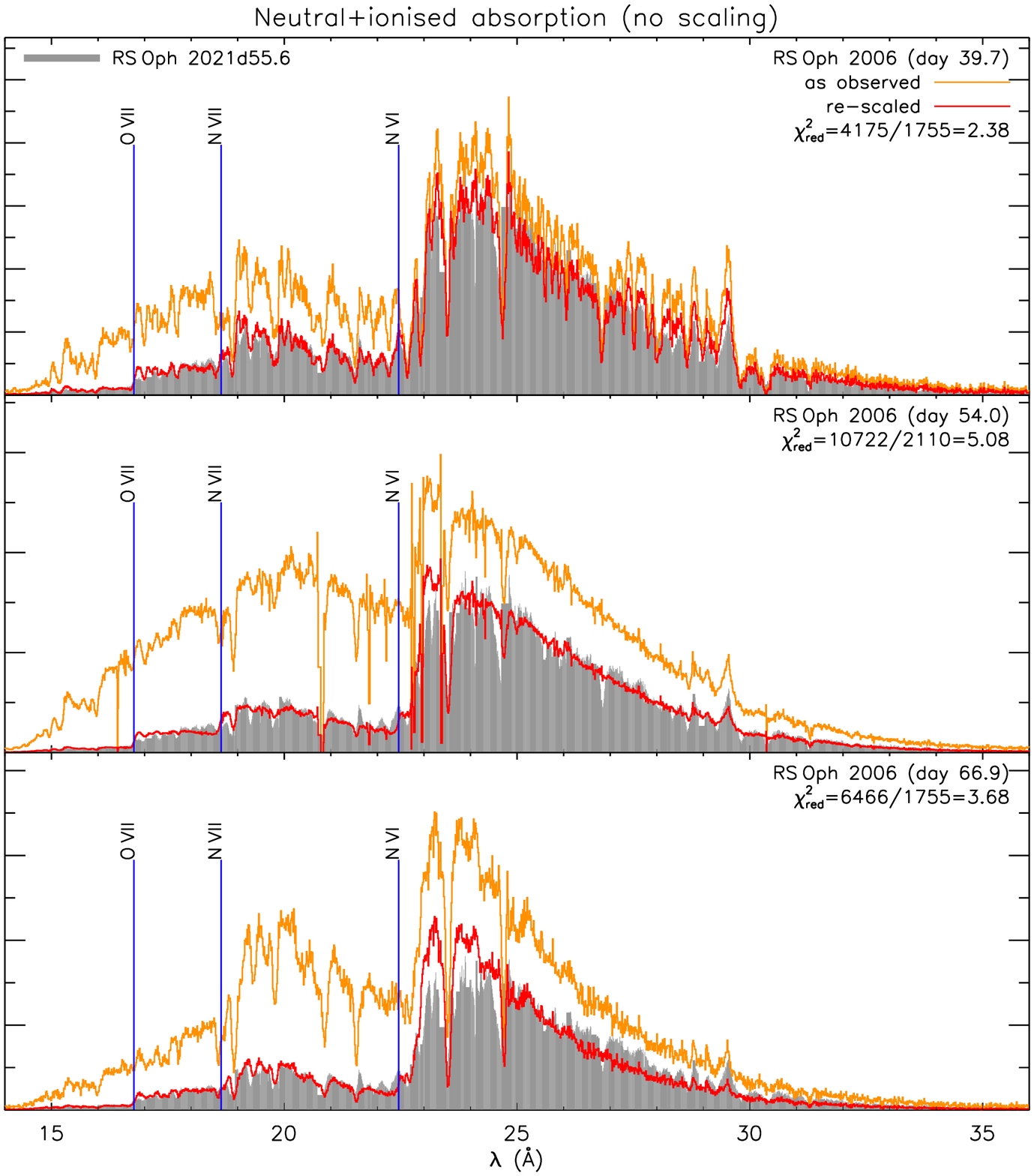}\includegraphics{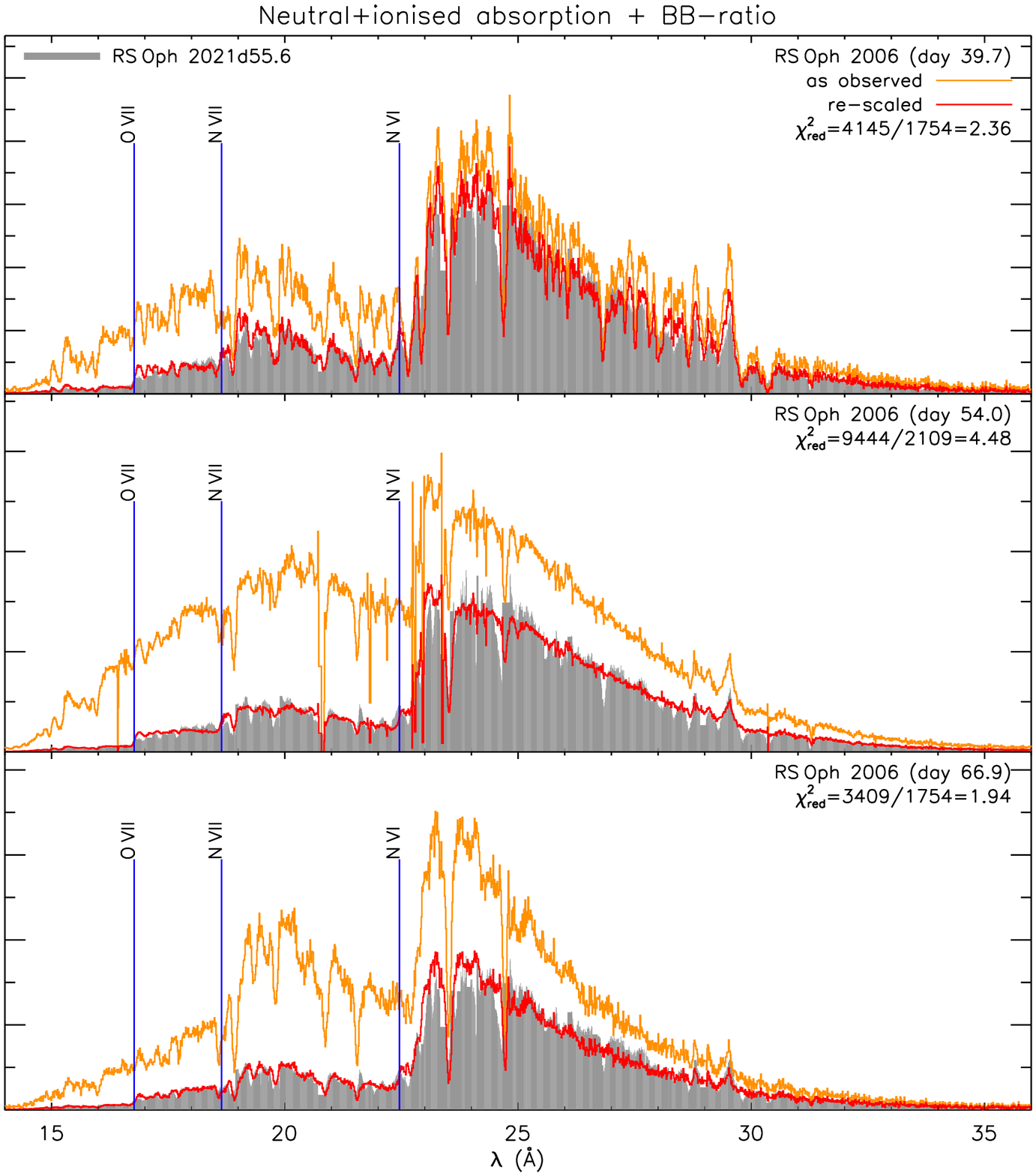}}
        \caption{\label{fig:xmmspec55}\xmm\ RGS1 spectrum obtained on 2021d55.6
        (grey shadings) compared to the 2006 spectra obtained on 2006d39.7
        (top), 2006d54 (middle), and 2006d66.9 (bottom). The orange lines
        represent the unscaled (brighter) 2006 spectra as observed, while the
        respective red lines are the results of applying an absorption
        model to the observed 2006 spectra assuming the absorption parameters
        given in Table~\ref{tab:absmodels}: $\Delta N_{\rm H}$ of an absorption
        model of only neutral absorbers with modification of the depth of the
        O\,{\sc i} absorption edge at 22.8\,\AA\ and the parameters of three absorption
        edges of highly ionised ions with values of optical depths and column
        density. A scaling factor was not applied. In the left panel, the
        result from scaling only with an absorption model is shown, while in
        the right panel, additional scaling with blackbody ratio has been applied,
        thus assuming a change in temperature between the spectra.
        The parameters of these and more models are given in Table~\ref{tab:absmodels}.
}
\end{figure*}

\subsubsection{Results}
\label{sect:spectra}

%The RGS spectrum for day 2021d55.6 is shown with grey shadings in photon
%flux units in Fig.~\ref{fig:xmmspec55} in comparison to selected scaled
%or unscaled other grating spectra of RS\,Oph as labelled in the respective
%legends. The 2006 SSS spectra taken on days 39.7, 54 and 66.9
%\citep{nessrsoph,nelson07} are used to visually test downscaling mechanisms described
%in \S\ref{sect:analysis:spec} that could explain the lower count rates during the
%2021 compared to the 2006 outburst.\\

\begin{figure}[!ht]
\resizebox{\hsize}{!}{\includegraphics{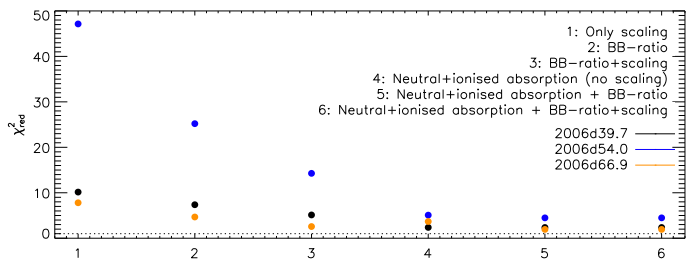}}
        \caption{\label{fig:chi}Reduced values of $\chi^2$ for the six
        down-scaling models listed in Table~\ref{tab:absmodels}. The dotted
        line in the bottom represents $\chi^2_{\rm red}=1$ (=ideal fit).
        Down-scaling
        by a wavelength-independent constant factor (number 1) shows the poorest performance,
        while the largest improvement is seen between combinations 3
        (scaled BB-ratio) and 4 (pure absorption without scaling). The
        down-scaling from day 2006d66.9 (orange symbols) behaves slightly
        differently, yielding a higher $\chi^2$ value for model 4 than for models
        3 and 5, and temperature changes are thus detectable (see text
        \S\ref{sect:spectra}).}
\end{figure}

\begin{figure}[!ht]
\resizebox{\hsize}{!}{\includegraphics{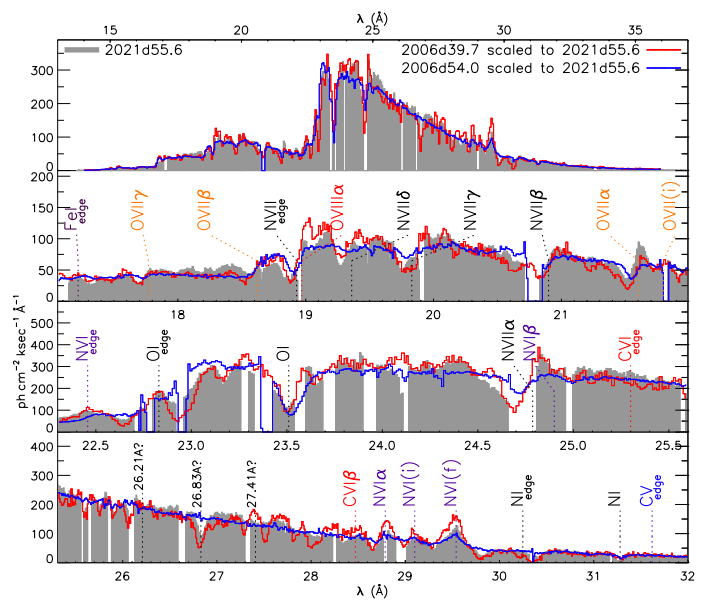}}
  \caption{\label{fig:cmpxmmspec55}\xmm\ RGS1 spectrum taken on 2021d55.6
        compared to 2006 spectra scaled with the pure absorption model (left
        column in Fig.~\ref{fig:xmmspec55}). The lower panels show details, and
        although the 2021d55.6 spectrum has a lot of gaps, one can see that
        the same-epoch 2006 spectrum (blue) contains much weaker absorption
        lines, while on the sub-\AA\ level, there is a greater resemblance
        to the 2006d39.7 spectrum.
        }
\end{figure}

\begin{figure}[!ht]
\resizebox{\hsize}{!}{\includegraphics{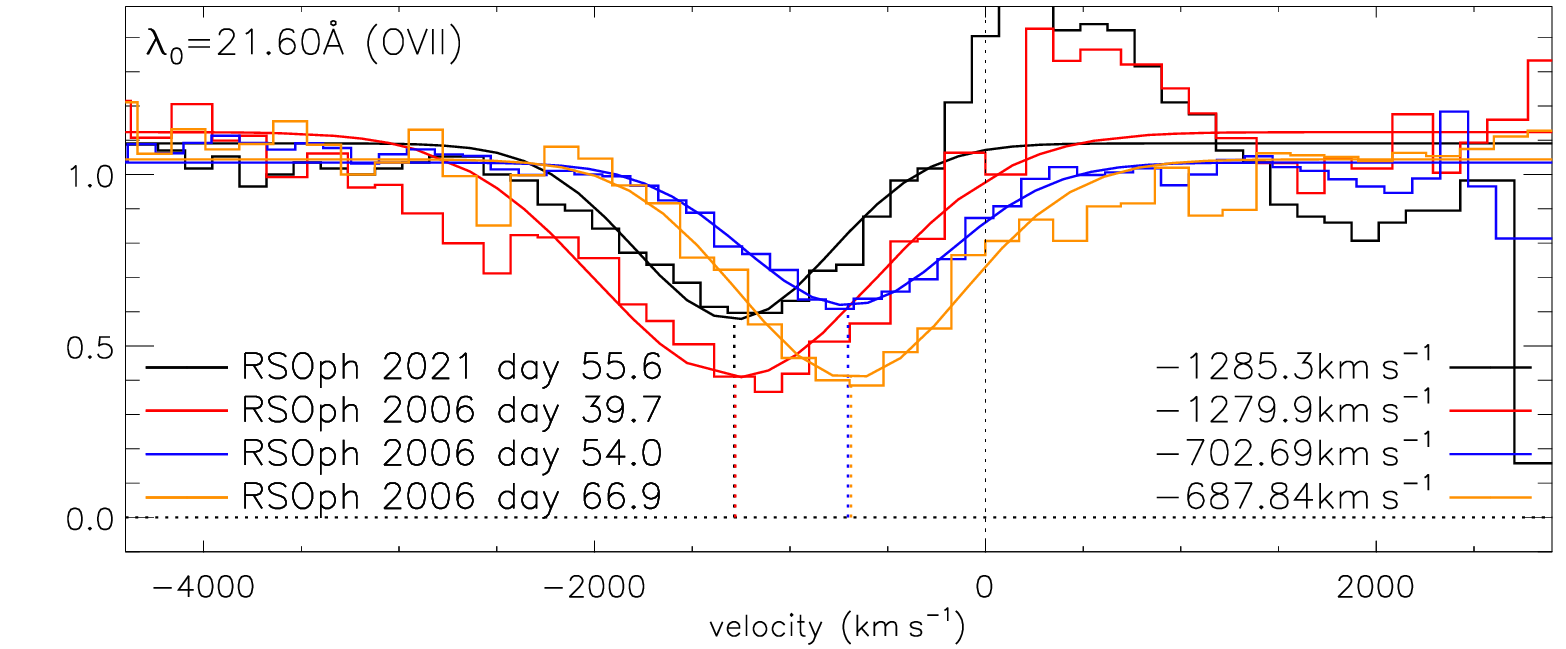}}
\resizebox{\hsize}{!}{\includegraphics{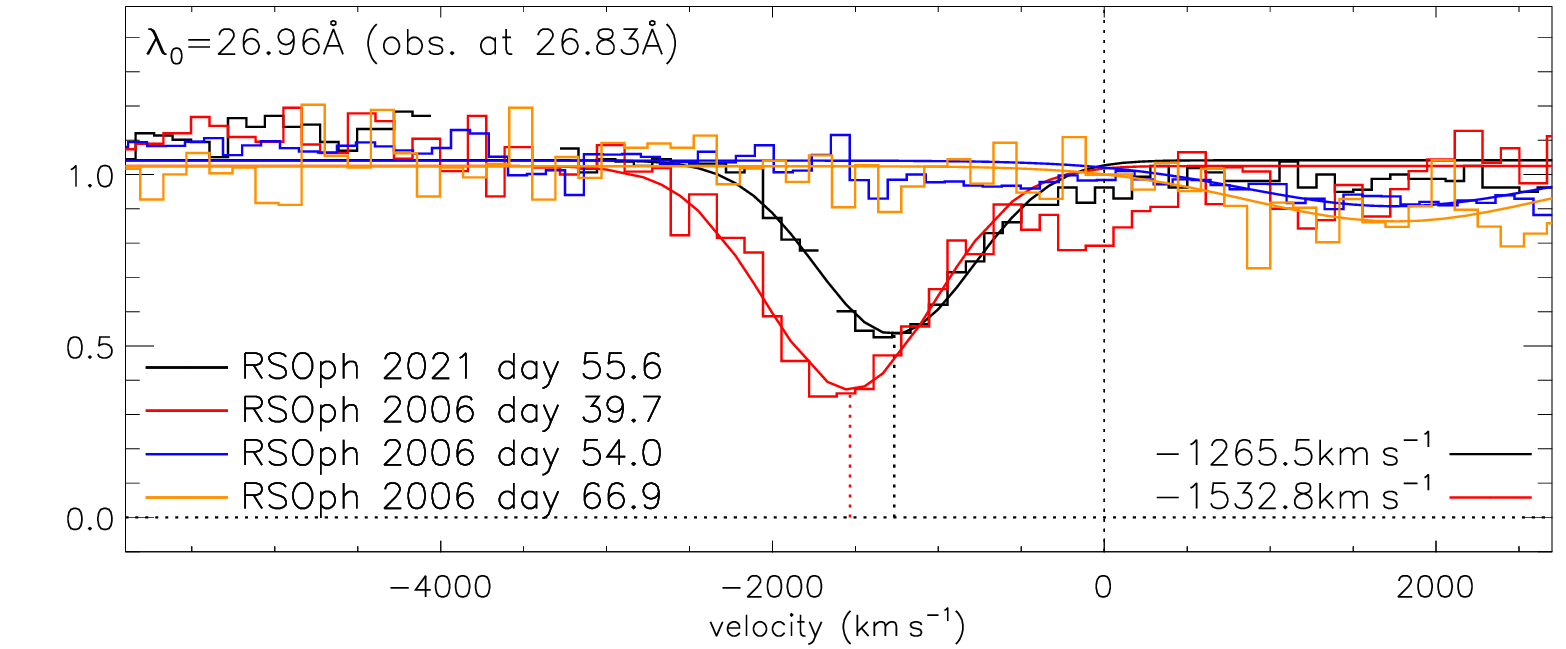}}
        \caption{\label{fig:OVII}Comparison in velocity space of
line profiles of the O\,{\sc vii} line at rest wavelength 21.6\,\AA\ (top)
and an unknown absorption line observed at 26.83\,\AA\ (bottom)
for the 2021d55.6 observation and the 2006d39.7, 2006d54, and 2006d66.9
observations. Assuming a similar blueshift for the unidentified line, we
adopt a rest wavelength of 26.96\,\AA.
Data are shown in histogram style and Gaussian fits to the
absorption lines with solid lines. Values of best-fit blueshift are
given in the bottom right legend.
Observed spectra and models are normalised to the respective median
values of the plot range.}
\end{figure}

The RGS spectrum for day 2021d55.6 is shown with grey shadings in photon
flux units in Fig.~\ref{fig:xmmspec55} in comparison to the bright SSS
continuum spectra taken on days 2006d39.7, 2006d54, and 2006d66.9
using orange and red lines for the original spectra and the down-scaled
spectra, respectively. We experimented with various ways to down-scale
the 2006 spectra to match the day 2021d55.6 spectrum as described
in \S\ref{sect:analysis:spec}, namely any combination of: (1) constant
scaling factor (\S\ref{sect:analysis:scal}), (2) multiplication
by the ratio of two different blackbody models (probing temperature
changes, \S\ref{sect:analysis:bbscal}), and (3) multiplication with
an absorption model (probing higher absorption in 2021 compared to 2006,
\S\ref{sect:analysis:absscal}).
For the ratio of two blackbody models, we fixed the first temperature
to $5\times 10^5$\,K given the degeneracy that for any temperature
value, a second temperature value can be found yielding the same
ratio curve (see \S\ref{sect:analysis:bbscal}).
As on days 2006d54 and 2006d66.9 the O\,{\sc i} absorption edge
was considerably reduced compared to 2006d39.7 \citep{nessrsoph},
we allow changes in the depth of the O\,{\sc i} edge when fitting
the neutral absorption model.\\

We present graphical representations of the results of two selected
approaches with the red curves in Fig.~\ref{fig:xmmspec55}: In the
left column, we show the results of assuming pure photoelectric absorption
by neutral and ionised material, and in the right column
with additional emission-spectrum temperature changes, both without scaling.
The agreement between the scaled 2006 and 2021d55.6 is remarkable, lending
significant credibility to our novel approach and demonstrating its ability to provide useful information within
a certain scope: For days 2006d39.7 and 2006d54, the agreement of the
unscaled absorption model is already excellent without the need
for the additional BB ratio. For day 2006d66.9, the pure absorption
model (bottom left) clearly leaves room for improvement, which can
be achieved by the additional assumption of a temperature change
(bottom right).\\

%The best-fit parameters are given in Table~\ref{tab:absmodel}.
%While an absorption model applied to an additive model only
%allows positive values of column densities, in our approach, a negative column
%means that it was lower in 2021 than in 2006.\\

\begin{table}[ht]
\begin{flushleft}
\renewcommand{\arraystretch}{1.1}
\caption{\label{tab:absmodels}Parameters of down-scaling 2006 spectra to the
        2021d55.6 spectrum. A graphical illustration of $\chi^2_{\rm red}$
        can be seen in Fig.~\ref{fig:chi}.}
\begin{tabular}{llll}
\hfill 2006:& d39.7 & d54 & d66.9\\
\hline
\multicolumn{4}{l}{Only scaling}\\
Scaling Factor & 0.60 & 0.34 & 0.40 \\
$\chi^2_{\rm red}$ & {\it 10.15} & {\it 47.14} & {\it 7.80} \\
\hline
\multicolumn{4}{l}{BB-ratio}\\
$T_{\rm eff}(1)/10^5$K & 5.00 & 5.00 & 5.00 \\
$T_{\rm eff}(2)/10^5$K & 5.21 & 5.43 & 5.38 \\
Scaling Factor & 1.00 & 1.00 & 1.00 \\
$\chi^2_{\rm red}$ & {\it 7.37} & {\it 25.20} & {\it 4.67} \\
\hline
\multicolumn{4}{l}{BB-ratio+scaling}\\
$T_{\rm eff}(1)/10^5$K & 5.61 & 4.60 & 4.24 \\
$T_{\rm eff}(2)/10^5$K & 6.50 & 5.60 & 5.15 \\
Scaling Factor & 2.89 & 4.45 & 5.61 \\
$\chi^2_{\rm red}$ & {\it 5.12} & {\it 14.27} & {\it 2.57} \\
\hline
\multicolumn{4}{l}{Neutral+ionised absorption (no scaling); Fig.~\ref{fig:xmmspec55} left}\\
$\Delta N_{\rm H}/10^{21}$cm$^{-2}$ & 0.34 & 0.68 & 0.76 \\
$\Delta$ O\,{\sc i} edge & -0.12 & 3.28 & 2.39 \\
$\Delta N_{\rm X}$ (O\,{\sc vii})/$10^{18}$cm$^{-2}$ & 5.00 & 5.60 & 4.90 \\
$\Delta N_{\rm X}$ (N\,{\sc vii})/$10^{18}$cm$^{-2}$ & 6.50 & 6.70 & 2.30 \\
$\Delta N_{\rm X}$ (N\,{\sc vi})/$10^{18}$cm$^{-2}$ & 2.30 & 2.60 & 2.80 \\
Scaling Factor & 1.00 & 1.00 & 1.00 \\
$\chi^2_{\rm red}$ & {\it 2.38} & {\it 5.08} & {\it 3.68} \\
\hline
\multicolumn{4}{l}{Neutral+ionised absorption + BB-ratio; Fig.~\ref{fig:xmmspec55} right}\\
$\Delta N_{\rm H}/10^{21}$cm$^{-2}$ & 0.42 & 0.35 & 0.01 \\
$\Delta$ O\,{\sc i} edge & -0.08 & 5.19 & 44.9 \\
$\Delta N_{\rm X}$ (O\,{\sc vii})/$10^{18}$cm$^{-2}$ & 5.10 & 5.00 & 3.50 \\
$\Delta N_{\rm X}$ (N\,{\sc vii})/$10^{18}$cm$^{-2}$ & 6.80 & 5.60 & -0.20 \\
$\Delta N_{\rm X}$ (N\,{\sc vi})/$10^{18}$cm$^{-2}$ & 2.50 & 2.10 & 2.30 \\
$T_{\rm eff}(1)/10^5$K & 6.54 & 5.46 & 4.90 \\
$T_{\rm eff}(2)/10^5$K & 6.50 & 5.60 & 5.15 \\
Scaling Factor & 1.00 & 1.00 & 1.00 \\
$\chi^2_{\rm red}$ & {\it 2.36} & {\it 4.48} & {\it 1.94} \\
\hline
\multicolumn{4}{l}{Neutral+ionised absorption + BB-ratio+scaling}\\
$\Delta N_{\rm H}/10^{21}$cm$^{-2}$ & 0.41 & 0.35 & 0.01 \\
$\Delta$ O\,{\sc i} edge & -0.08 & 5.20 & 57.9 \\
$\Delta N_{\rm X}$ (O\,{\sc vii})/$10^{18}$cm$^{-2}$ & 5.20 & 4.90 & 3.50 \\
$\Delta N_{\rm X}$ (N\,{\sc vii})/$10^{18}$cm$^{-2}$ & 6.80 & 5.60 & -0.21 \\
$\Delta N_{\rm X}$ (N\,{\sc vi})/$10^{18}$cm$^{-2}$ & 2.50 & 2.10 & 2.30 \\
$T_{\rm eff}(1)/10^5$K & 6.54 & 5.46 & 4.90 \\
$T_{\rm eff}(2)/10^5$K & 6.50 & 5.60 & 5.15 \\
Scaling Factor & 1.00 & 1.00 & 1.02 \\
$\chi^2_{\rm red}$ & {\it 2.36} & {\it 4.48} & {\it 1.94} \\
\hline
\end{tabular}
\end{flushleft}
\end{table}

Table~\ref{tab:absmodels} lists best-fit parameters testing various
combinations of scaling, temperature-changes (BB ratio), and absorption.
The $\chi^2_{\rm red}$ values in the last row of each respective section
indicate how well the scaled 2006 spectra in each column reproduce the
2021d55.6 spectrum. A graphical illustration of these values is shown in
Fig.~\ref{fig:chi}. The different combinations are ordered in
Table~\ref{tab:absmodels} and Fig.~\ref{fig:chi} (see legend) by degree of
complexity. Pure scaling factor and scaled and unscaled
BB-ratio fits yield rather poor agreement in
all cases, while the pure absorption model alone, without any scaling
(fourth section in Table~\ref{tab:absmodels}; see also left panel of
Fig.~\ref{fig:xmmspec55}), leads to much better agreement with the
data on 2021d55.6, except for the scaling from day 2006d66.9.
Adding a scaled or unscaled BB ratio (fifth section in
Table~\ref{tab:absmodels}) only leads to improvement for day
2006d66.9, while for day 2006d39.7 there is no
improvement. The combined absorption plus BB-ratio model
for day 2006d66.9 works better than a pure BB-ratio model.
No neutral $\Delta N_{\rm H}$ is needed, while the improvement
is driven by the ionisation edges, where the N\,{\sc vii} edge
is found to be shallower (negative value) on day 2021d55.6 than
on day 2006d66.9. Additional scaling (last section in
Table~\ref{tab:absmodels}) leads to only small improvements in fit
in all three cases.
For day 2006d66.9, we can therefore conclude that a temperature change
must be responsible for the difference in brightness, while
different absorption edges of ionised material are also important.
We note that we cannot quantify the temperature difference as the
fitted temperature depends on the choice of the first temperature.
Also, the ratio of temperatures is not meaningful as it also depends
on the choice of the first temperature (see Fig.~\ref{fig:bbtest}).
All we can conclude with this method is that the
effective temperature of the underlying source in the
2006d66.9 spectrum was higher than on day 2021d55.6, although we are not able to quantify by how much.\\

These results lead us to the conclusion that absorption is the dominant factor explaining the lower fluxes in 2021 compared to 2006, while a temperature change
is only detectable in the last 2006 observation, which was taken after the peak
of the SSS emission was reached (see Fig.~\ref{fig:mmlc}).\\

Another interesting result is that the agreement between day 2021d55.6 and
2006d39.9 is actually better than that with the same-epoch observation 2006d54.
While in 2006, the neutral oxygen column density decreased substantially by
day 54, possibly by continuous photoionisation \citep{nessrsoph}, the same
process leading to this reduction apparently did not take place during the
2021 outburst (or was slower).
%For the downscaling of the day 2006d54, we have to account for significant
%additional column densities of the ionisation edges of ions such as O\,{\sc vii}
%to get agreement with the day 2021d55.6 spectrum, and it is reasonable to
%expect the same differences between the depth of the respective absorption lines.
%The additional O\,{\sc vii} emission line on days 2006d39.7 and 2021d55.6
%may have been produced in additional material that causes absorption in the
%line of sight but emission away from the line of sight. On day 2006d54
%less such material may have been present leading to less absorption and fewer
%line emission.\\
Figure~\ref{fig:cmpxmmspec55} shows that also on the sub-\AA\ level,
2021d55.6 agrees better with 2006d39.7 than with the same-epoch
2006d54 spectrum, both yielding deeper absorption lines with more
similar blueshifts.
Unfortunately, due to the missing information from the RGS2, the day 2021d55.6
spectrum contains a lot of gaps, and few absorption line profiles qualify
for direct comparison with all three spectra.
In the top panel of Fig.~\ref{fig:OVII}, we show a comparison of the (normalised)
O\,{\sc vii} ($\lambda_0=21.6$\,\AA) line region in velocity space, again
showing much more
similarity between days 2021d55.6 and 2006d39.7. As discussed by
\cite{nessrsoph}, the early 2006 SSS spectra still contained P-Cyg-type
line profiles, while the emission line component disappeared in the later
observations. It can be seen in Fig.~\ref{fig:OVII} that an O\,{\sc vii}
emission line was clearly present on day 2021d55.6, while not on days
2006d54 and 2006d66.9.
Also, the observed blueshift was much lower on day 2006d54 than on both 2006d39.7 and 2021d55.6.
Another absorption feature suitable for comparison is an unknown absorption
feature observed at 26.83\,\AA. In the bottom panel of
Fig.~\ref{fig:OVII}, we show the profile assuming a rest wavelength of
26.93\,\AA\ (yielding about the same blueshift as for O\,{\sc vii} on
day 2021d55.6). This line has no emission line component, and it is only
present on days 2006.39.7 and 2021d55.6, again showing these two observations
are more similar than the same-epoch observation on day 2006d54. We note that the
blueshift for the 26.83-\AA\ line is different on days 2006.39.7 and 2021d55.6,
while for O\,{\sc vii} this blueshift is consistent on both days.
The day 2021d55.6 spectrum therefore resembles an absorbed version of the
2006d39.7 spectrum, suggesting that the opacity of the ejecta  to soft X-rays
may have evolved more slowly during the 2021 outburst.\\

\subsection{Spectral variations}
\label{sect:specvar}

\begin{figure}[!ht]
\resizebox{\hsize}{!}{\includegraphics{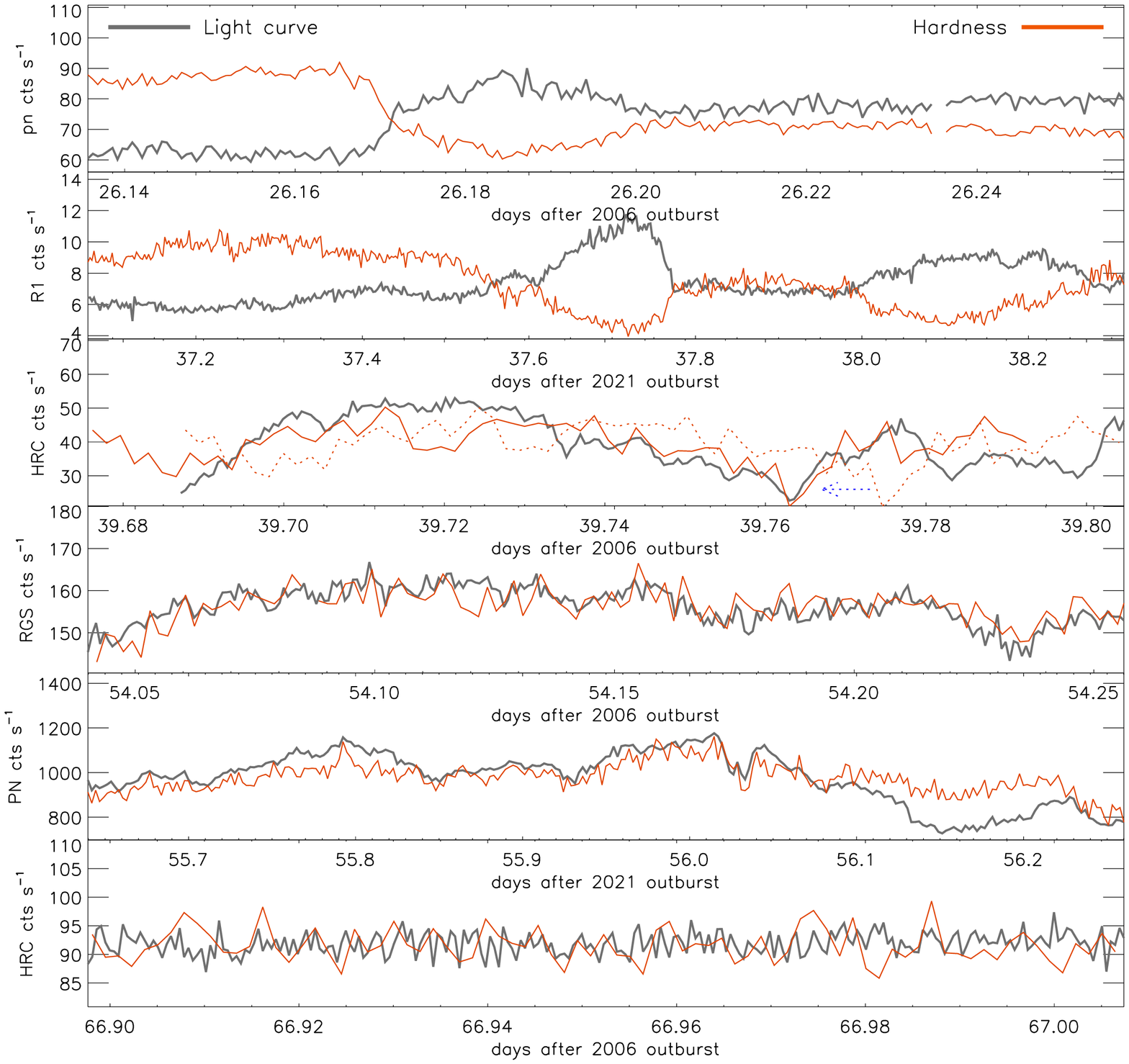}}
        \caption{\label{fig:cmphr}Comparisons of 2006 and 2021 light curves
        with the respective hardness-ratio light curves (see respective
        horizontal axes labels). In the third panel (2006d39.7), the
        observed hardness ratio light curve is shown with a dotted line while the
        solid line represents the hardness ratio light curve shifted by -1000s
        (see \citealt{nessrsoph,schoenrich07,nesspalermo}).
        No such lag is seen on days 2006d54 and 2021d55.6, although a small lag
        might be present. On days 2006d26.1 and 2021d37.1 (top two panels),
        hardness and brightness are anti-correlated, reflecting the appearance
        and disappearance of the complex soft emission component, while much less variability
is seen during the
        decline observed on day 2006d66.9 (bottom panel).
}
\end{figure}

\begin{figure*}[!ht]
\resizebox{\hsize}{!}{\includegraphics{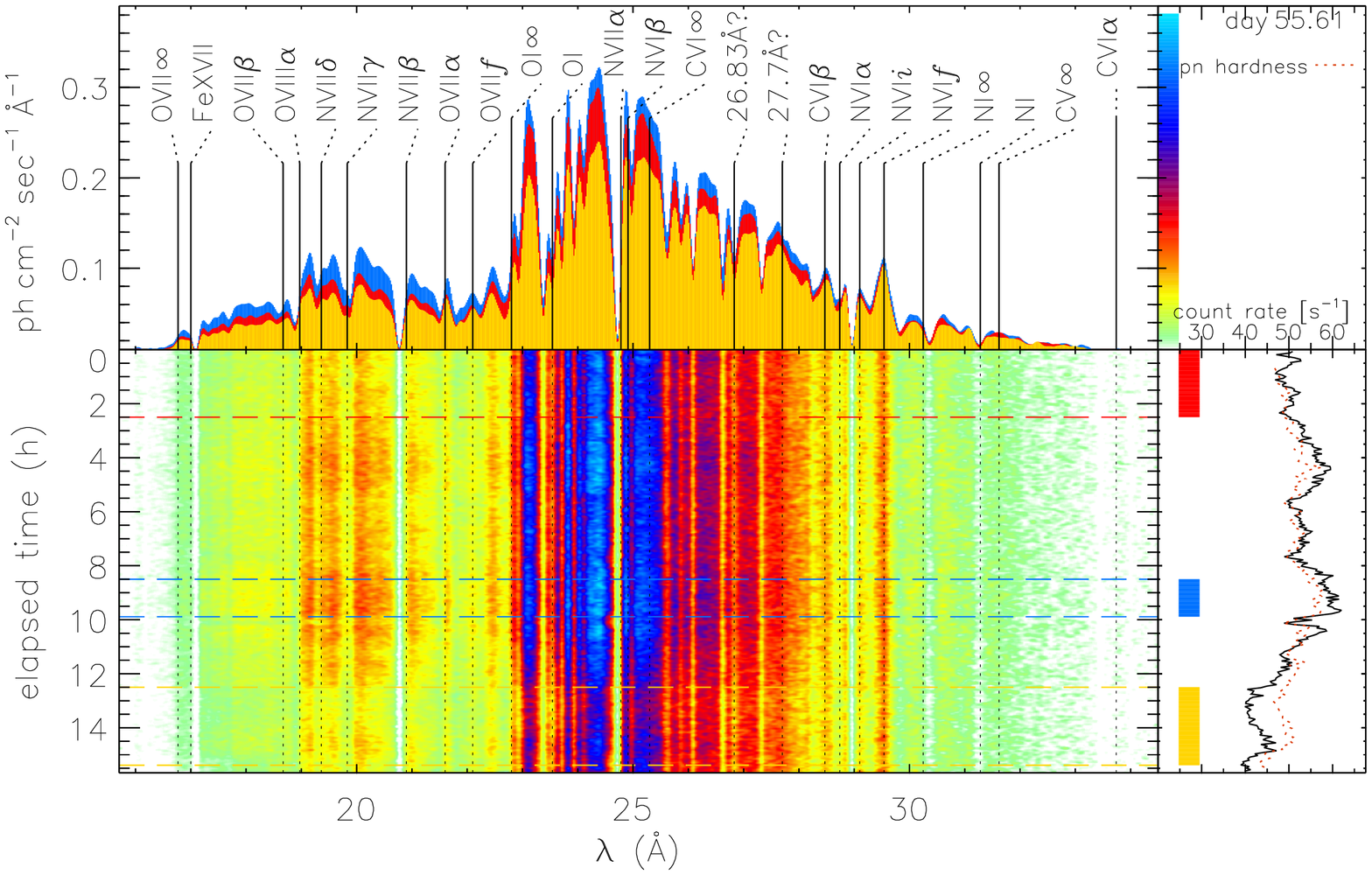}}
        \caption{\label{fig:smap55}Spectral time map based on 281 RGS spectra
extracted from adjacent 200s time intervals from the 2021d55.6 observation.
The main panel shows the
spectra with wavelength from left to right, time from top to bottom, and colours representing flux
following the (non-linear) bar in the top right panel along the vertical
flux axis. The dashed horizontal lines and shadings with the same colours in the
bottom right panel indicate time intervals from which the spectra shown with the
same colour shadings in the top left panel were extracted.
The brown dotted line in the light curve panel in the bottom right represents the hardness light curve extracted from the pn where hardness is scaled to fit into the graph while varying between values -0.580 and -0.397.
}
\end{figure*}

In Fig.~\ref{fig:xmmlc}, variability can be seen to correlate with variations
in the hardness ratio. Such variations may be due to variations in the
O\,{\sc i} column density that would cause variations in the depth of the
absorption edge at 540\,eV, the minimum energy needed to ionise O\,{\sc i} out of
the inner (K) shell.
To test this hypothesis, we extracted the hardness-ratio light curve from the pn data
split at the energy 540\,eV defining the hardness ratio as $HR=(H-S)/(H+S)$ with $H$ and $S$
being count
rates extracted from the respective energy bands $H=540-800$\,eV and $S=300-540$\,eV.
The hardness ratio varies between values of -0.580 and -0.397 on day 2021d55.6,
and for comparison with the brightness evolution, we have scaled the
hardness ratio curve to a curve with the same median value and dynamic range (thus
the same minimum and maximum values) and overplot the thus-scaled hardness curve in the
second panel of the right graph in Fig.~\ref{fig:xmmlc} with the brown curve.
Here, we see that brightness and hardness variations are directly correlated,
indicating that the depth of the O\,{\sc i} absorption edge has caused the
variations.\\

Figure~\ref{fig:cmphr}  shows hardness (separated at the O\,{\sc i}
ionisation edge at 540\,eV) and brightness light curves for all
SSS observations in the order of epoch, starting with the two
early SSS observations on days 2006d26.1 and 2021d37.1 until the post-peak
observation taken on day 2006d66.9. During the early-SSS observations,
hardness and brightness are strongly anti-correlated, reflecting that all
variability happened in the soft band while the hard band represented
only the more constant shock emission. We note that during these early SSS
observations, there was no SSS emission shortward of the O\,{\sc i}
absorption edge (Fig.~\ref{fig:xmmspec37}).\\

Changes of the O\,{\sc i} absorption edge proportional to brightness changes
were already found during the 2006 outburst on day 39.7 \citep{nessrsoph},
where the hardness evolution lagged 1000 seconds behind the brightness
evolution. \cite{nessrsoph,schoenrich07} interpreted this lag as owing to
density-dependent changes in the ionisation or recombination timescales.
In the fourth panel of Fig.~\ref{fig:cmphr}, it can be seen that on day 2006d54, variations
were almost exactly coincident with possibly a small lag of hardness
following brightness changes. Such a small lag may actually also be
present in the day 2021d55.6 observation (fifth panel). However, towards
the end of the day 55.6 observation, the correlation seems to break down,
where we remind the reader that the 35s period seems to be strongest;
see Fig.~\ref{fig:lmap}. After the peak of SSS emission on day
2006d66.9, we see no variability (bottom panel). If nuclear burning had
already turned off, this would indicate that the variability observed
in the other SSS light curve depends on active central burning.
A similar behaviour was seen in M31N 2008-12a by \cite{M3112a2016}
with the SSS light curve becoming smoother with the transition from
a burning white dwarf to a cooling white dwarf.\\

To visualise the spectral evolution in a continuous manner, we use the
concept of spectral time maps. The general concept of these spectral
time maps is described in \cite{basi}, for example.
In the spectral time map shown in Fig.~\ref{fig:smap55}, it can be seen that the
emission around 20\,\AA\ increased by more during the brighter phases than the
emission above $22.8$\,\AA\ (blue and red spectra in the top panel). The
variations therefore appear to be related to changes in the O\,{\sc i}
absorption edge: during brighter episodes, the O\,{\sc i} column density was
lower and thus more transparent to harder emission.
The coloured bars in Fig.~\ref{fig:xmmlc} mark the time intervals
from which spectra were extracted, representing selected activity
episodes that are shown in the top panel of Fig.~\ref{fig:smap55}.
During the last episode of lower count rates
(yellow spectrum), the count rate is not only lower at $\lambda<22.8$\,\AA\
but also above, and the origin of this reduction in flux is therefore different.\\

\section{Discussion}
\label{sect:disc}

We present a novel data-to-data scaling approach enabling us to compare
different data sets with minimal model assumptions. This approach does not
require a full description of all features, because those appearing in both spectra 
cancel out. The fact that we find both differences in absorption and
changes in intrinsic properties demonstrates the ability of this
approach to distinguish absorption from intrinsic changes.\\

This method can be applied to all SSS spectra extracted from different brightness
phases of the same system, testing whether fainter SSS emission was due to higher
absorption of different temperature or lower intrinsic luminosity. For example,
we applied this test to the high-flux and low-flux \xmm\ spectra of V2491\,Cyg, ObsID
0552270501 \citep{nessv2491}, clearly finding that a small temperature change can fully
account for the brightness change, even without re-scaling. A systematic
study of all SSS spectra would therefore likely be fruitful.

\subsection{Explaining the lower SSS flux in 2021}
\label{sect:disc:sss}

The 2006 spectra on days 39.7 and 54 can be brought into agreement with the
2021 spectrum on day 55.6 by down-scaling them with a relatively simple
absorption model involving cold and hot overlying material (top two panels
on the left of Fig.~\ref{fig:xmmspec55}).
It is worth noting that no additional scaling is required, and therefore the lower
total flux on day 2021d55.6 is consistent with the spectral changes expected
from more photoelectric absorption. The addition of changes of temperature
(top two panels in the right of Fig.~\ref{fig:xmmspec55})
only improves the agreement of the re-scaled 2006 spectra with the
2021d55.6 spectrum by a minimal amount. The fact that
the best-fit temperatures for blackbody scaling are so close to each other
supports absorption as the dominant factor for the reduced brightness in 2021.\\

%\begin{figure}[!ht]
%\resizebox{\hsize}{!}{\includegraphics{fe10}}
%        \caption{\label{fig:fe10}Line fluxes of [Fe\,{\sc x}] at 6375\,\AA\
%       for the outbursts of 1985 \citep{anupama1989}, 2006 \citep{munari2007},
%       and the 2021 outburst \citep{page2022}.
%}
%\end{figure}

The intrinsic source was then equally bright in both outbursts. This is supported
by the observation that the line flux light curves of the coronal [Fe\,{\sc x}] emission line
(6375\,\AA) shown in figure~14 by \cite{page2022} are similar in the two
outbursts. The [Fe\,{\sc x}] line is a forbidden line and as such is optically
thin; it is therefore more capable of escaping much denser plasma without re-absorption
or scattering than resonance
lines. We are therefore seeing the sum of [Fe\,{\sc x}] emission from all
regions, and line-of-sight effects are irrelevant as local inhomogeneities
balance out.
Radiative decays of forbidden lines occur on long timescales, and the
higher the plasma density (and therefore the collision rate), the higher the number
of excited states that are
depopulated by collisions rather than the radiative de-excitations that
produce the line photons. While a higher plasma density increases the
population of the ionisation stage such as Fe\,{\sc x}, the same higher
density reduces the production efficiency of forbidden lines such as the
[Fe\,{\sc x}] line at 6375\,\AA. Meanwhile, photoionisation can increase the
Fe\,{\sc x} number density without decreasing the strength of forbidden lines,
and the intensity of coronal lines such as [Fe\,{\sc x}] are therefore indicators
of the local presence of a strong central ionising radiation source, even
if this latter does not materialise as observed soft-X-ray emission.
[Fe\,{\sc x}] line measurements catch photoionising radiation emitted in
all directions, which is the SSS intensity averaged over
$4\pi$. Meanwhile, soft-X-ray observations can be compromised by
line-of-sight effects, such as angle-dependent emission and/or absorption
in different directions.
We therefore consider [Fe\,{\sc x}] measurements to be an important source
of complementary information to soft-X-ray observations.
%, and
%in Fig.~\ref{fig:fe10}, we reproduce the 2006 and 2021 [Fe\,{\sc x}]
%line fluxes shown in figure~14 in \cite{page2022}, adding
%[Fe\,{\sc x}] measurements by \cite{anupama1989} taken during the 1985
%outburst.
From figure~14 in \cite{page2022} and figure 5 in \cite{munari2022},
we highlight
\begin{itemize}
\item The steep increase in [Fe\,{\sc x}] intensity between days $\sim 23$ and $30$
indicates
that the SSS phase already started a few days before soft-X-ray emission was
detected.
\item No high-amplitude variations in the [Fe\,{\sc x}] line.
\item The [Fe\,{\sc x}] line fluxes are similar in 2006 and 2021.
\end{itemize}

These differences support the conclusion of intrinsically the same SSS
emission, while emission and absorption are not spherically
symmetric, which means that what
we see in soft X-rays depends on the viewing angle. The \swift/XRT analysis carried out
by \cite{page2022} does not obviously support significant differences in
absorption, while the grating spectra analysed with our approach of
data scaling provide strong evidence for absorption being the dominant
factor explaining the lower observed flux in 2021 compared to 2006.
We interpret these differences as due to the lower spectral resolution
of the \swift/XRT data and the limitations on the analysis methods that are possible
under the circumstances of a high degree of statistical photon redistribution:
\cite{page2022} fitted blackbody and atmosphere models
to the XRT spectra taken at similar epochs in 2006 and 2021 and compared
the resulting parameters of source and absorption models. While these
source models fit the 2021 XRT spectra well (see figure 9 in \citealt{page2022})
both blackbody and atmosphere models are only approximations and have
never reproduced X-ray high-resolution grating spectra. The good agreement
between these models and the \swift/XRT spectra is therefore mostly
owing to the low resolution of the XRT `washing out' narrow features that would be
difficult to reproduce with models. At the low spectral resolution, any
deficiencies in the source models will be washed out and can easily be
compensated for by the absorption model; this freedom reduces the
sensitivity to differences in absorption. By replacing the source model
with the 2006 data, this freedom is being removed, which substantially
increases the sensitivity of the differences in absorption.\\

As discussed by \cite{page2022}, the finding that absorption is the reason for
the lower soft-X-ray flux in 2021 compared to 2006 is a challenge to the geometry.
One could argue that the different orbital phases in 2006 (0.26) and
2021 (0.72) may imply a different portion of phase-aligned absorbing structures
such as the accretion stream or overlying envelope of the companion star
to be located within the line of sight. However, \cite{page2022} argue that
the 1985 {\it EXOSAT} soft-X-ray observations were taken during a similar orbital
phase (0.32) to in 2006 while even lower soft-X-ray fluxes were seen;
however, it should be noted that the temporal coverage of the 1985 outburst in the X-ray was far sparser, and gross variability may again have played a part in the observed difference in flux.
Phase-aligned large-scale structures are therefore less
likely to be accountable for the differences in absorption.
However, we note that a phase difference of 0.26-0.32=0.06 translates to a duration of
27.2 days assuming an orbital period of $453.6\,\pm\,0.4$\,days
\citep{brandi09} which is of order of the same timescale as the
evolution of the nova outburst. We also note that the uncertainty of the orbit
of 0.4 days accumulates to an uncertainty in phase of 2\%
after 17 orbits between 1985 and 2006. The overlap in phase is therefore marginal
with a range of 0.28 - 0.34 in 1985 and 0.24 - 0.28 in 2006.
Comparison of the 2006 and 2021
[Fe\,{\sc x}] line fluxes suggests that the overall SSS brightness was similar in these
two outbursts, and we searched for similar diagnostics for the 1985 outburst.
[Fe\,{\sc x}] line fluxes are given by \cite{anupama1989}, but even if
reversing their de-reddening, they are not consistent with the 2006 and 2021
fluxes. We attribute these inconsistencies to high uncertainties owing to the
the non-linear behaviour complicating the conversion from photographic density
integrated over a line profile into a linear flux value. Unfortunately, we found no
publication covering the 1985 event where the spectra were recorded  with other
than photographic plates.
We inspected the Asiago photographic plate archive finding
38 spectra of RS Oph in 1985 \citep[cf][]{rosino1987}.
From these original plates, we found that in most
observations, the [Fe\,{\sc x}] line suffers saturation in its core, making
an accurate flux calibration impossible during the most interesting brightest
time interval of the evolution. We then scaled the [Fe\,{\sc x}] fluxes
against He\,{\sc i,} finding that the 1985 spectra agree relatively well with
the 2006 and 2021 data.\\

So, we cannot clarify whether or not the large differences are phase-dependent.
An assumption of the surrounding material to be randomly inhomogeneous
(independent of phase) has the weakness of not being testable but it could lead to
small changes in viewing angle to cause large changes in the SSS intensity
and spectral shape. This could also explain the high-amplitude variations during
the first $\sim 3$ weeks of the SSS phase in 2006 that \cite{osborne11} concluded
to be caused, at least partly, by clumpy ejecta.
While we have no grating observations covering the low- and high states of
these variations, it appears possible that spectra extracted from low-flux
episodes can be reproduced by application of similarly simple absorption
models to the spectra from the high-flux episodes. This would need to be
tested with a long grating observation covering low- and high-flux episodes.
The \xmm\ proposal under which this was attempted led to the observation on
day 2021d37.1; however, the early variability phase was not observed as
expected, and confirming this hypothesis will require a high-risk (but also highly
rewarding) target-of-opportunity observation of a future nova in outburst.

\subsection{Explaining post-peak differences}

The 2006d66.9 spectrum that was obtained after the peak of
SSS emission (see Fig.~\ref{fig:mmlc}) cannot reproduce the 2021d55.6
spectrum with only absorption (bottom left panel in Fig.~\ref{fig:xmmspec55}),
and the addition of a temperature change is required (bottom right panel
in Fig.~\ref{fig:xmmspec55}). The resulting temperature parameters indicate
that the effective temperature at the pseudo-photosphere was higher on
day 2006d66.9 than it was on
2021d55.6. As nuclear burning may have already turned off after the
peak of SSS emission was reached, one might naively expect the temperature
to start decreasing. However, our observations are not driven by the
temperature of the central source but by the radiation transport behaviour of
the higher layers, and we may be seeing deeper into the outflow.
%if the outer layers are starting to fall in, as a consequence of reduced
%radiation pressure from the fading central radiation source, then they will
%heat via compression.\\

\subsection{Explaining the different spectral structures}

Figure~\ref{fig:cmpxmmspec55} demonstrates that the
shape and structure of the SSS spectrum on 2021d55.6 more closely resembles the 2006d39.7 spectrum than the same-epoch 2006d54 spectrum.
Specifically, the following items were found to be common to 2006d39.7 and
2021d55.6 while different from 2006d54:
\begin{itemize}
\item The O\,{\sc i} absorption edge was much deeper in 2006d39.7 and 2021d55.6.
\item The 2006d39.7 and 2021d55.6 spectra contain more and deeper absorption lines (Fig.~\ref{fig:cmpxmmspec55}).
\item The absorption lines of O\,{\sc vii} (Fig.~\ref{fig:OVII}), for example,
        were blueshifted by
        $\sim 1200$\,km\,s$^{-1}$ on days 2006d39.7 and 2021d55.6 but only by
        $\sim 700$\,km\,s$^{-1}$ on day 2006d54.
\end{itemize}

%Meanwhile, the short-term variability was correlated with hardness ratio
%variability in all three observations, however, with a 1000-s delay on
%day 2006d39.7.\\

%From Fig.~\ref{fig:mmlc} one can see that in 2021, SSS emission became
%bright some 10 days later than in 2006, and one may conclude that the
%evolution of the opacity of the ejecta within our line of of sight was
%then also delayed by 10 days.
%Since the end of SSS emission was not delayed, there was no systematic
%shift of the SSS phase, but the early opacity evolution of the ejecta
%along our line of sight
%seems to have been slower in 2021, and the conditions on day 2006d39.7
%might have been reached $\sim 10$ days later in 2021. That could have
%led to same-epoch data not being consistent with each other:
%On day 2021d37.1, we still saw highly absorbed SSS emission similar
%to day 2006d26.1, while on day 2021d55.6, the opacity was closer
%to that on day 2006d39.7 than on day 2006d54.\\

%But how did that affect the shape and structure of the SSS spectra?
The deeper O\,{\sc i} absorption edge on day 2021d55.6 compared to
day 2006d54 can be interpreted as the absorbing plasma having been
hotter in 2006, leading to oxygen being ionised to higher ionisation
stages or even being fully ionised (discussed in more detail
in \S\ref{sect:disc:oedge}). Also the presence of more absorption
lines on day 2021d55.6 compared to day 2006d54 can be interpreted
as a higher temperature on day 2006d54 causing absorption lines from low-ionisation
elements to disappear, while the fully ionised (higher) fraction of the material
in the line of sight produces no absorption lines. Meanwhile, the observed expansion velocity
was lower on day 2006d54 than on day 2021d55.6, indicating that
slower-moving material was hotter, while the plasma containing
lower-temperature spectral features moved at higher velocities.\\

Concentrating on the 2006 evolution for a moment, the evolution
from day 2006d39.7 to 2006d54 can be understood as follows:
\begin{itemize}
\item The nova ejecta decelerate while transferring their kinetic
        energy to the stellar wind of the giant companion star,
        and this kinetic energy feeds the observed shock emission.
\item While the outflow was decelerated, it was also heated and
        (collisionally) ionised, which is a natural consequence of the evolution
        of shock systems generated by the interaction of the nova
        ejecta with the pre-existing stellar wind.
\item The higher degree of ionisation on day 2006d54 compared to
        the earlier day 2006d39.7 made the absorbing plasma more
        transparent leading to the higher observed flux on day 2006d54.
\end{itemize}

In this picture, originally lower-temperature ejecta with a higher
opacity were moving at $\sim 1200$\,km\,s$^{-1}$ before entering the
regions of the stellar wind where they gradually decelerated to
$\sim 700$\,km\,s$^{-1}$ while increasing the degree of ionisation
and thus transparency to soft X-rays.
Meanwhile, in 2021, this process was less
efficient, as on day 2021d55.6 we see more absorption, no reduction
in velocity, and no reduction in  the number of low-temperature absorption
lines. If this is due to the different viewing angles in 2006 and 2021,
the interactions between nova ejecta and stellar
wind are not homogeneous in all directions, which could be caused by an
inhomogeneous density distribution of the wind of the stellar companion for example.
In this picture, the line of sight coincided in 2021 with lower-density
regions of the stellar wind, and the plasma could
expand more freely in that direction without detectable deceleration
and heating via collisions.\\

\subsection{Variations in the O\,{\sc i} edge depth}
\label{sect:disc:oedge}

In this section, we discuss possible interpretations for the observed variations
in the depth of the O\,{\sc i} K-shell absorption edge at 540\,eV (22.8\,\AA),
which implies variations in the O\,{\sc i} column density. These could either be
caused by ionisation of O\,{\sc i} to higher ionisation stages or a more
radical change of the entire oxygen abundance.\\

In the 2006 observations of RS\,Oph, \cite{nessrsoph,schoenrich07,nesspalermo}
found variations in the absorption edge to correlate with source brightness,
and concluded that ionising O\,{\sc i} is responsible for the variable
transparency for hard emission. However, there are a few potential issues
with this interpretation.
With the brightness variations as the trigger for changes in the
        ionisation balance, independent intrinsic brightness variations
        have to be assumed. In this scenario, the brightness variations cannot be a
        consequence of variable absorption, at least not fully. Meanwhile,
        correlated hardness and brightness changes are interpreted as a consequence
        of variable absorption, and cause and effect are in conflict.\\

If O\,{\sc i} is ionised, the number density of higher ionisation stages
        increases. While \cite{nessrsoph} identified absorption lines from
        O\,{\sc ii} and higher ionisation stages of oxygen (see their
        Table~3), we find from the {\tt ionabs} model that there will
        be ionisation edges at higher energies that we do not observe.
        The only way to avoid additional edges would be to ionise all the
        way to O\,{\sc vii} or higher.\\

For the absorbing material to respond to the brightness changes,
        some time is needed. While there should be enough time for ionisation
        between days 2006d39.7 and 2006d54, more rapid variations of
        the O\,{\sc i} column density are observed within all SSS observations
        (Fig.~\ref{fig:cmphr}). There must be some response time leading
        to a lag between brightness and hardness changes, but we only observe
        such a lag in the observation on day 2006d39.7.
        This lag could therefore be explained as a response time to intrinsic
        brightness changes, but the lack of lag in all the other observations
        is then difficult to explain.\\

Interestingly, the lag duration of 1000s corresponds to a light travel
        spatial scale of
        $\sim 2$\,AU; however, the absorber must be in the same line of sight
        as the continuum source, and if it is the source of ionisation,
        any brightness changes would be delayed by the same light travel time.\\

If oxygen is photoionised by an increased level of radiation intensity,
        then also hydrogen, helium, and other elements with lower ionisation
        energies than 540\,eV should be ionised. However, the depth of the
        O\,{\sc i} edge was observed to vary more strongly than the low-energy
        slope of the continuum, indicating that the O\,{\sc i} edge variations
        are decoupled from changes of $N_{\rm H}$.
        The likelihood of a photon ionising oxygen or hydrogen decreases
        exponentially with the difference between photon energy and
        ionisation energy, and 540 eV photons ionise oxygen much more efficiently
        than hydrogen.
        An unabsorbed SSS spectrum approximated by a blackbody with
        $T_{\rm eff}=5\times 10^5$\,K
        peaks at 200\,eV with 540\,eV in the Wien tail with $\sim 6$\%
        intensity relative to the peak. The same 6\% intensity level is
        encountered in the Rayleigh-Jeans tail at 54.7\,eV, not far from the 13.6 eV
        ionisation energy of hydrogen. It is therefore not conceivable for oxygen
to be photoionised while hydrogen is not. A way out may be that
        circumstellar hydrogen, helium, and so on is already fully ionised, while
        some neutral oxygen is initially still present in the local circumstellar
        material.\\

On the other hand, to produce the Raman-scattered O\,{\sc vi} line
        observed by \cite{munari2022}, a great amount of neutral gas is
        required (at least neutral hydrogen), and the evolution of this
        line suggests that the global amount of neutral hydrogen remained fairly
        constant during the SSS phase.\\ 
	
	An alternative explanation for changes in the O\,{\sc i} column density could
be found by reducing the overall oxygen abundance in general by condensation of
oxygen into dust grains. This would reduce the number of oxygen atoms that can
absorb photons above 540\,eV. While some of the issues with the ionisation hypothesis
could be avoided, there are also arguments against dust formation.\\

Dust grains may lead to other observable features in the
        X-ray spectra. These features can be studied with the SPEX model as
        has been demonstrated by \cite{pinto12}. We have not performed such
        modelling as it is beyond the scope of this work. We can therefore
        at present not decide whether or not dust formation is consistent with the
        SSS spectra.\\

As the O\,{\sc i} absorption edge not only decreased but
        also increased on short timescales, a dissociation mechanism is also needed.
        It appears unlikely that the number of oxygen atoms
locked in dust can vary as quickly as we observe the depth of the
O\,{\sc i} edge to vary. A possible explanation of the strictly correlated
hardness and brightness variations may be that clumps with oxygen already
locked in grains moved into and out of the line of sight and all variability
that correlates with hardness would then be due to variations in obscuration
with a reduced O\,{\sc i} column density. This would also be a way out for
                the ionisation hypothesis.\\

There is no evidence from either excess infrared emission or
breaks in the optical light curve of any large-scale dust formation following the
outburst \citep{evans88}. While there will be some destruction of dust in the
pre-existing stellar wind of the companion star, some of it survives
\citep{evans2007}, but whether or not that is sufficient to explain changes in the
O\,{\sc i} edge on the observed timescales would require studies beyond the
scope of this work.\\

Cyclic creation and destruction of dust seed nuclei were found
        by \cite{oxygenflaring} for the DQ Hercules-like nova V5668 Sagittarii
        (2015), but again the timescales on which this can lead to
        destruction and replenishment of oxygen may be slower than the
	fast changes in O\,{\sc i} that we observed.\\

Photon-induced condensation and dissociation would actually lead to an
        anti-correlation of hardness ratio and brightness as a higher radiation
        intensity would lead to increased dissociation and thus a deeper
        O\,{\sc i} edge.\\

We cannot fully rule out nor confirm either mechanism. We lean towards
the ionisation hypothesis.
%In the bottom panel of Fig.~\ref{fig:cmphr}, no variability is seen in the
%post-SSS-peak observation on day 2006d66.9. If at that time, nuclear
%burning has switched off, the variability in all other observations
%might depend on nuclear burning to be active.
%{\bf TBD convection?}

\section{Summary and conclusions}
\label{sect:concl}

The results from this study can be summarised as follows:
\begin{itemize}
\item The significantly lower emission during the SSS phase in 2021 reported by \cite{page2022} can be explained by absorption from cold (neutral) and hot (ionised) material in the line of sight. This can be demonstrated by application of a relatively simple absorption model to the bright 2006 SSS spectra taken on days 39.7 and 54, which leads to almost perfect agreement with the RGS1 spectrum taken on day 55.6 after the 2021 optical peak.
The direct inference is that the intrinsic SSS emission, and thus energy production,
was the same in 2006 and 2021. This is consistent with similar [Fe\,{\sc x}] fluxes observed in 2006 and 2021.
\item The post-peak SSS spectrum on day 2006d66.9 can also be rescaled to match
        the 2021d55.6 spectrum, but in addition to absorption, we also detect a 
        higher effective temperature in 2006.d66.9. We conclude that after the SSS
        peak, we can see hotter layers after the pseudo-photosphere has receded
        further inside.
\item The 2021d55.6 spectrum more closely resembles the 2006d39.7 spectrum than the same-epoch
        2006d54 spectrum:
        the scaling by a simple absorption model works better; the O\,{\sc i} absorption edge is deeper than on day 2006d54; the spectrum on day 2006d54 contains fewer absorption lines; similar P-Cyg-like absorption line profiles are observed, while on day 2006d54 no emission line components were present,
        and the blueshift on days 2006d39.7 and 2021d55.6 was $\sim 1200$\,km\,s$^{-1}$
        while on day 2006d54 we find only $\sim 700$\,km\,s$^{-1}$.
        From our discussion, we conclude from these differences that the
        deceleration of the nova ejecta was not the same in all directions
        owing to an inhomogeneous density distribution of the wind of the
        stellar companion. The total amount of dissipated energy must have
        been the same in 2006 and 2021 because the intensity of bremsstrahlung
        emission radiated in all directions was the same
        \citep{page2022,orio2022}. Meanwhile, the small portion of
        the nova ejecta propagating in the direction of the line of sight
        underwent a different degree of deceleration and heating between days
        $\sim 40$ and $60,$ depending on the density of the stellar wind in the
        respective directions probed in 2006 and 2021.
        %We conclude that those ejecta components producing the observed absorption
        %lines were gradually slowed by dissipating kinetic energy in the wind of
        %the stellar companion. This process has taken place more efficiently in 2006
        %than in 2021 resulting in higher velocities on day 2006d39.7 than on
        %on day 2006d54. During this process, the temperature of the absorbing plasma
        %increased via collisional ionisation as a side effect of the interactions
        %between ejecta and stellar wind of the companion star leading to a reduced
        %number of absorption lines by day 2006d54. Meanwhile, in 2021, the absorbing
        %plasma had a higher column density and was not slowed and not heated.
        %If this would be a viewing angle effect, the interactions with the stellar wind are not equally efficient in all directions.

\item The X-ray light curves taken during the bright SSS phase (days 2006d39.7,
        2006d54, and day 2021d55.6) are mildly variable, and the hardness ratio is correlated with the brightness changes. The hardness changes are caused by variability in the column density of neutral oxygen.
\item The hardness changes lag behind the brightness changes by 1000s only on day
2006d39.7 but by at most marginal lags on days 2006d54 and 2021d55.6.
\item Attributing variations in the O\,{\sc i} absorption edge to ionisation may have some issues.
      Ionising O\,{\sc i} to O\,{\sc ii-v} would lead to other features that we do not observe.
      A reduction in the O\,{\sc i} edge without seeing other edges at
                higher energies could be achieved by immediate ionisation to
                O\,{\sc vii} or higher; however, this should take more time
                than the 1000s lag observed on day 2006d39.7 and appears
                inconsistent with the instantaneous hardness and brightness variations
                seen in all other observations.
      If oxygen is photoionised, elements with lower ionisation
                energies such as hydrogen and helium must also be photoionised,
                and fully ionised material would be fully transparent.
      An alternative to ionisation may be condensation and/or dissociation of
                oxygen into dust grains but this is inconsistent with infrared observations.
 Further studies are required to investigate other potential inconsistencies, such as needing to see dust features
                in the X-ray spectra (testable with SPEX models) or timescales for
                condensation and dissociation of dust compared to observed timescales
                of variations.
\item In the new 2021 data, we find the 35s period with a lower duty cycle
        compared to the brighter day 2006d54 light curve.
        This suggests that the period is more difficult to detect in obscured data, which is consistent with its origin towards the  surface of the white dwarf.
\end{itemize}

Our novel approach to rescaling a grating spectrum with (a) a single scaling
factor, (b) an absorption model, or (c) the ratio of two blackbody models proves
to be a powerful tool for understanding brightness changes more generally. This approach
can be applied to revisit all variable SSS observations in the archive, and as an
example, we already find that the deep dip in the SSS light curve
of V2491\,Cyg \citep{nessv2491} can be explained by a small temperature change.

\begin{acknowledgements}
A.P. Beardmore, K.L. Page acknowledge support from the UK Space Agency.
M.J. Darnley receives funding from UKRI grant number ST/S505559/1.
A. Dobrotka and J. Magdolen were supported by the Slovak grant VEGA 1/0408/20, and by the European Regional Development Fund, project No. ITMS2014+: 313011W085.
S. Starrfield gratefully acknowledges partial support from NSF and NASA grants to ASU.
\end{acknowledgements}

\bibliographystyle{aa}
\bibliography{cn,jn,rsoph,astron}

\appendix

\section{Appendix}
\label{appendix}

\subsection{Correction for pileup in RGS}
\label{append:pileup}

Pileup is an effect caused by two (or more photons) arriving at the same position within the same readout time, and instead of registering two photons with their respective energies, a single photon with the sum of their energies is recorded. This is an issue for particularly bright sources with photon arrival rates that are faster than the readout time of the recording detector. Photon pileup is usually not an issue for the RGS because the source photons are distributed over a large area on the detector following the dispersion relation,  and therefore soft photons are recorded further away from the zeroth order than higher-energy photons. However, SSS belong to the rare class of sources for which even the RGS suffers pile up at those positions of the recording CCD chip array where the dispersed spectrum is brightest.

Fortunately, owing to the extreme softness of the source, the photons are not lost. At each chip position, dispersed photons can be identified from their photon energies in the {\tt pi} column of the events file to be consistent with the wavelengths computed from their angular distance from the zero-th order. Discrepancies in {\tt pi} energy and wavelength from the dispersion relation can be used to filter out background events. If a {\tt pi} energy yields twice the value of the corresponding wavelength, there are two possibilities, either the photon belongs to the second dispersion order, or two photons of the same energy have arrived at the same time, thus pile up.

The default {\tt rgsproc} pipeline assumes no pile-up and registers all photons with twice the {\tt pi} energies than the corresponding wavelength values in a separate second-order spectrum assuming the wavelength values to be the correct ones. This spectrum is a mix of piled photons and second dispersion order but for soft sources such as SSS, the number of second-order photons at high energies is small, and the second-order spectrum essentially consists of half the number of all piled photons at half their wavelength values.
For these photons, the values in the {\tt pi} column can be corrected to half their value, and as two photons had originally arrived, all photons for which the {\tt pi} energy was corrected need to be duplicated, preserving energy, which means additional rows added to the events file.\\

\begin{figure}[!ht]
\resizebox{\hsize}{!}{\includegraphics{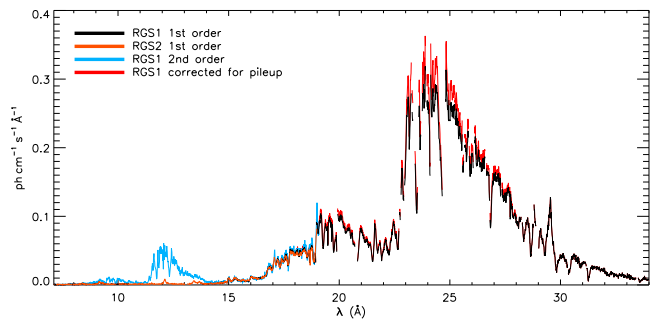}}
        \caption{\label{fig:pileup}Illustration of RGS pileup correction for ObsID 0841930901 (see \S\ref{append:pileup}. The Black spectrum shows the uncorrected RGS1 first-order spectrum, and the uncorrected RGS1 second order spectrum (light blue), a fainter clone of first-order spectrum can be seen at half the wavelengths. The second order spectrum captures the pileup photons that need to be moved into the first order, which then leads to the read curve, the corrected RGS1 spectrum.
}
\end{figure}

%The draw-back is that these photons cannot be distinguished from the second-order dispersion. While the process is different, and photons of a given energy are dispersed with twice the dispersion angle, the outcome is the same, and therefore photons at a given chip position contain twice the {\tt pi} column value.\LEt{Please check that I have retained your intended meaning.} The {\tt rgsproc} reduction chain thus records all piled photons together with the second-order dispersed photons in the second-order spectrum. A correction is therefore only possible if the second-order spectrum at half the wavelengths of the soft continuum spectrum is negligibly weak.
This is illustrated in Fig.~\ref{fig:pileup} where the RGS1 and RGS2 first-order spectra are shown with black and orange, respectively. For the RGS2, only the range up to 19\,\AA\ is shown as it has not operated correctly at longer wavelengths. The RGS1 second-order spectrum, shown with light blue, can be seen to be much brighter between 8\,\AA\ and 14\,\AA\ than the RGS2 first-order spectrum\footnote{In this wavelength range, there is no RGS1 first-order spectrum because of a chip gap.}. All emission in excess of the RGS2 first-order spectrum in this range is therefore due to photon pile up, and the correction procedure adds a small amount of emission from the Ne\,{\sc x} line at 24\,\AA.\\

Specifically, the steps to correct for photon pileup in the RGS are:
\begin{itemize}
 \item Regular {\tt rgsproc} run
 \item In the events files {\tt P0841930901R1S002EVENLI0000.FIT} and {\tt P0841930901R2S004EVENLI0000.FIT}, identify all events satisfying the following conditions for their column values {\tt m\_lambda} (in \AA) and {\tt pi} (in eV):
   \begin{itemize}
      \item $18 < {\tt m\_lambda} < 28$
      \item Ratio of column values {\tt pi}/(12398.5/{\tt m\_lambda})$=2\pm0.2$
   \end{itemize}
   where 12398.5/{\tt m\_lambda} represents the conversion from wavelength in \AA\ to photon energy in eV.
 \item For each of these events, change the value of {\tt pi} to half its value.
 \item To conserve the total energy, duplicate all these modified events.
 \item Based on the modified events file, produce all follow-up products with {\tt rgsproc entrystage=4:spectra}.
\end{itemize}

These manipulations can be performed with any software allowing fits file editing; we used IDL.

\subsection{Aliases from Telemetry Drops in the pn light curve ObsID 0841930901}
\label{append:TM}

\begin{figure}[!ht]
\resizebox{\hsize}{!}{\includegraphics{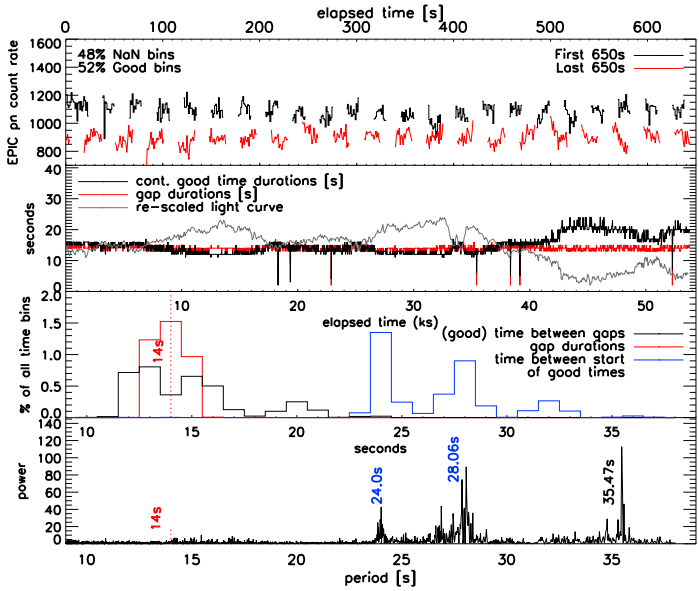}}
\caption[dbin]{\label{fig:dbin}Illustration of telemetry losses in the soft
(0.2-1.5\,keV) pn light curve taken on day 2021d55.6. {\bf Top}: First and last
600s time segments of the light curve showing regular gaps between time
intervals taking data. During the last 600s segment,
with a lower count rate, the duration of continuous good times are slightly
longer, illustrating the dependence of losses with count rate.
{\bf Second panel}: time evolution of durations of continuous good time intervals
(black) and gaps without data (red) in comparison to the (re-scaled) count
rates (grey). The durations of continuous data-taking were longer while the
source was fainter.
{\bf Third panel}: Histogram distributions of durations of good-times
(black), gaps (red), and times between starting good-time intervals (blue).
The durations of gaps vary within very small margins,
independent of count rate, between $13-15$\,s while the durations of good
times vary between $\sim 10-20$\,s (dependent on count rate, see panel above).
The distribution of times between start times of data-taking intervals
cluster at 24s and 28s, and these peaks introduce an alias to period studies
as seen in the {\bf bottom panel} where the periodogram contains peaks at
these periods.
}
\end{figure}

Another issue are telemetry losses owing to full memory buffer. For very
bright sources, not all events can be downlinked before the onboard
memory is full. This has occurred in the second observation (day 55.6) for
the pn in a quasi-periodic manner
as illustrated in the top panel of Fig.~\ref{fig:dbin}. Close inspection
shows that during time segments with lower count rate, the short time
intervals during which data could be obtained were slightly longer while
the gaps between these data chunks remained roughly the same. The evolution
of the durations of gaps and times during which data were downlinked (good
time intervals) are illustrated in the second panel, clearly showing
an anti-correlation of count rate (grey) and duration of continuous good times
(black). Therefore, when the count rate is lower, the capability to take
continuous data is higher. However, once a gap
is reached,  its duration varies only very little, that is, by $13-15$\,s.
The distribution of durations of good-times (black), gaps (red), and times
between start times of good-time-intervals (blue) are shown
in the third panel, where the (count rate independent) distribution of
gaps is strongly peaked at $14\,\pm\,1$ seconds. The bottom panel shows the
periodogram that does not contain a peak at the 14s gap duration but two
peaks at $\sim 24$s and $\sim 28$s, that coincide with the peaks in the panel
above showing the distribution of differences of start times of good-time
intervals. These peaks in the periodogram are therefore not related to the source
but are induced by the telemetry gaps, while the 35.47s period is not related
to the telemetry dropouts and can therefore be associated to the source
(see \S\ref{sect:variability}).

\end{document}